\documentclass[aps,prb,twocolumn,groupedaddress,showpacs,superscriptaddress,amssymb,amsmath,noeprint]{revtex4-2}
\usepackage{graphicx}
\usepackage[english]{babel}
\usepackage{dcolumn}
\usepackage{bm}
\usepackage{hyperref}
\usepackage{cleveref}
\usepackage{multirow}
\usepackage{braket}
\hypersetup{
    colorlinks=true,
    linkcolor=blue,
    urlcolor=blue,
    citecolor=blue
}

\usepackage{xcolor}
\usepackage{comment}

\newcommand{\tr}{\mathrm{Tr}}

\definecolor{bkac}{rgb}{1,0,0.5647}

\definecolor{bka}{rgb}{0.0588,0.3216,0.7294}

\begin{document}
\title{Understanding synchronization between quantum self-sustained oscillators through coherence generation}

\author{Mohit Kumar}
\email{mohit.kumar@students.iiserpune.ac.in}
\affiliation{Department of Physics, Indian Institute of Science Education and Research Pune, Dr. Homi Bhabha Road, Ward No. 8, NCL Colony, Pashan, Pune, Maharashtra 411008, India}

\author{Bijay Kumar Agarwalla}
\email{bijay@iiserpune.ac.in}
\affiliation{Department of Physics, Indian Institute of Science Education and Research Pune, Dr. Homi Bhabha Road, Ward No. 8, NCL Colony, Pashan, Pune, Maharashtra 411008, India}

\date{\today} 

\begin{abstract}
Understanding the origin of phase synchronization between quantum self-sustained oscillators has garnered significant interest in recent years. In this work, we study phase synchronization in three settings: between two continuous-variable oscillators, between two arbitrary quantum spins, and within a hybrid setup involving a spin and an oscillator. We derive a simple and general condition on the elements of the joint density matrix that must be satisfied for them to contribute to the relative phase distribution. In particular, we identify the subset of coherence elements in the joint density matrix that serve as key resources for enabling quantum phase synchronization. Our theory is validated against the previously proposed interaction models known to induce synchronization between the self-sustained oscillators. Moreover, our approach offers valuable insights into the relationship between phase synchronization and various information-theoretic measures.
\end{abstract}

\maketitle

\section{Introduction}\label{sec:Introduction}

Synchronization refers to the adjustment of rhythms between interacting oscillatory systems \cite{Pikovsky}. This is a ubiquitous phenomenon observed across a wide range of disciplines, including biology, engineering, chemistry, and physics \cite{Sync_sci_tech,Sync_complex_network,Kuramoto}. Notable examples include the collective flashing of fireflies and the synchronized oscillation of metronomes. In classical systems, the emergence of synchronization is well understood through extensive studies of paradigmatic models such as the Kuramoto model \cite{Kuramoto_model_1,Kuramoto_model_2,Kuromoto_RMP_2005,kuramoto1984chemical} and the Van der Pol oscillator \cite{Class_VdP_1,Class_VdP_2,Class_VdP_3}. The exploration of this phenomenon in the quantum domain has garnered significant attention over the past decade \cite{Qubit2008,Qubit2009,Giorgi-1,Giorgi-2,1-QHO,1-Spin,Dissipative_Sensing,Strong_nonlinear,Macroscopic}, driven both by fundamental questions, as well as the promise of novel applications in emerging quantum technologies.

Quantum synchronization has been studied in a wide range of physical systems, including continuous-variable (CV) setups such as quantum Van der Pol oscillators \cite{1-QHO,2-QHO,Lee-1,Topological_sync_vdp,ENMZF_PRR_2020,Rvdp,Critical_response} and bistable oscillators \cite{JLA_PRR_2020}. It has also been explored in systems without clear classical analogues, such as discrete spin systems \cite{1-Spin,2-Spin,BlockadeBruder,Non_reciprocal_spin,Half_integer} and hybrid configurations involving both oscillators and spins \cite{Measures_Sai}. Recent advancement in quantum technologies have further accelerated this field, enabling the experimental realization of quantum synchronization across various experimental platforms, such as cold atoms \cite{Cold_atom_sync}, optomechanical setup \cite{Optomechanical_sync}, the IBM quantum computer \cite{IBM_sync}, micromechanical oscillators \cite{micromechanical_sync}, and nuclear spin systems \cite{Nuclear_spin_sync}.

Unlike in classical systems, quantum synchronization is fundamentally governed by features unique to the quantum regime, such as coherence, quantum correlations, and quantum noise. To characterize synchronization in quantum systems, several measures have been proposed, including the ones derived from phase distribution \cite{Spin_correlations,Anharmonic_blockade,micromasers_sync}, as well as information-theoretic quantities such as mutual information \cite{Mutual_info} and relative entropy of synchronization \cite{Measures_Sai}.

In this work, we focused on understanding the origin of quantum synchronization through the lens of relative phase distribution, considering two mutually interacting self-sustained oscillators (SSOs), realized in continuous-variable, discrete-spin, and hybrid oscillator-spin configurations. In particular, we investigate the role of quantum coherence in enabling phase synchronization between such self-sustained oscillating systems. While it is known that for a single driven SSO, all coherences between energy eigenstates serve as a resource for synchronization \cite{Resources1SSO}, the connection between coherence and synchronization in the case of two mutually coupled SSOs remains less well understood. Here, we demonstrate that only a specific subset of coherence elements of the joint density matrix, in the eigenbasis of the bare system Hamiltonian, contributes to phase locking, while coherences outside these subsets are completely irrelevant to phase synchronization. 

The above perspective not only provides insights into the microscopic origin of quantum synchronization but also offers 
a pathway to engineer interactions and dissipative mechanisms that can selectively generate relevant coherences. To substantiate this point, we analyze various forms of coherent and dissipative couplings in this work. Our findings also offer valuable insights into the applicability of different proposed measures of quantum synchronization.

The paper is organized as follows: In Sec.~\ref{sec:Sync-VdP}, we begin by analyzing the continuous-variable setup. Using the Wigner quasi-probability distribution, we demonstrate that a specific subset of density matrix elements—those satisfying the excitation conservation relation—are responsible for phase locking. Furthermore, we investigate the generation of these elements under various forms of interactions that have previously been shown to induce synchronization.  In Sec.~\ref{sec:sync-spin} and Sec.~\ref{sec:sync-hybrid}, we generalize this result for discrete spin systems and for a hybrid configuration consisting of a CV system and a spin. In Sec.~\ref{sec:coh_corr}, we examine the relationship between coherence, correlation, and synchronization. Additionally, we analyze various information-theoretic measures of quantum synchronization and shed light on their applicability. Finally, we summarize our results in Sec.~\ref{sec:summary}. Additional technical details are provided in the appendices.

\section{Synchronization of Two continuous-variable Oscillators} 
\label{sec:Sync-VdP}
In this section, we first start by investigating phase synchronization between two quantum CV oscillators. A paradigmatic model in this case is the quantum Van der Pol (VdP) oscillator \cite{Lee-1,1-QHO}, which is described as a quantum harmonic oscillator subjected to linear gain and nonlinear loss. The interplay between these two processes leads to the emergence of a characteristic limit-cycle behavior. The dynamics of a single quantum VdP oscillator is governed by the Lindblad quantum master equation,
\begin{equation}
   \dot{\rho} = -i \left[ \omega_0 a^\dagger a, \rho \right] + \gamma_g \mathcal{D}[a^\dagger]\rho  + \gamma_l \mathcal{D}[a^2]\rho,
   \label{eq:QME-VdP}
\end{equation}
where $\omega_0$ is the natural frequency of the oscillator and $\mathcal{D}$ denotes the Lindblad dissipator, defined as $\mathcal{D}[O]\rho = O\rho O^\dagger - \frac{1}{2} \{ O^\dagger O, \rho \}$. Here, $\gamma_g$ and $\gamma_l$ represent the rates for linear gain and nonlinear loss, respectively. In the long-time limit, the steady state solution of Eq.~\eqref{eq:QME-VdP} is diagonal in the Fock basis and exhibits an annular ring-shaped Wigner function \cite{Lee-1,1-QHO}, with its maximum displaced from the origin -- an indicator of limit-cycle formation. The circular symmetry of this distribution further signifies the existence of a free phase.

In the following, we first identify the key resource responsible for synchronization between two such continuous-variable SSOs, and then analyze various forms of mutual interaction that give rise to phase locking between them.

\subsection{Relative Phase Distribution and The Excitation Relation}
\label{subsec:relative-phase}
In this subsection, we derive the condition under which relative phase locking emerges between two continuous-variable SSOs. Let $\rho$ denote the joint density matrix of two quantum harmonic oscillators. Since the complete information about their phase correlations is encoded in the joint state $\rho$, we begin by analyzing the joint Wigner distribution, defined as \cite{Lee-2, carmichael}
\begin{align}
   & W(x_1, p_1, x_2, p_2) = \frac{1}{\pi^2} \int_{-\infty}^{\infty} \!  dy_1 \int_{-\infty}^{\infty} dy_2 \, \, e^{2i (p_1 y_1 + p_2 y_2)}  \nonumber \\
 & \quad \qquad \times \langle x_1 - y_1, x_2 - y_2|\rho|x_1+ y_1, x_2 + y_2\rangle.
 \label{eq:two-mode Wigner}
\end{align}
The relative phase is defined as $\phi=(\phi_1-\phi_2)$, where $\phi_1$ and $\phi_2$ denote the individual phases of the two oscillators. To extract information about the relative phase, we apply the following coordinate transformation to the Wigner function
\begin{equation*}
(x_1,p_1,x_2,p_2)\to (r_1,\phi_1,r_2,\phi_2)\to(r_1,r_2,\phi,\phi_2),
\end{equation*}
where the first transformation maps the cartesian coordinates to polar coordinates, and the second transformation replaces the phase variable $\phi_1$ with $(\phi+\phi_2)$. These change of variables yields a joint probability distribution over $r_1,r_2,\phi$, and $\phi_2$. By integrating out all variables except $\phi$, we obtain the marginal distribution $P_{w}(\phi)$ for the relative phase $\phi$. The subscript $w$ indicates that this distribution is derived from the Wigner function. Expanding the joint density matrix $\rho$ in Eq.~\eqref{eq:two-mode Wigner} in the joint Fock basis $\{\ket{m_1,m_2}\}$ of the two harmonic oscillators, we can express the relative phase distribution $P_{w}(\phi)$ in the following form (see Appendix \ref{sec: AppendixA} for the details of the derivation)
\begin{widetext}
\begin{align}
    P_{w}(\phi) &= \sum_{m_1, n_1, m_2, n_2=0}^{\infty} \!\! \langle m_1, m_2 |\rho| n_1, n_2 \rangle \int^{\infty}_0 dr_1\  r_1 \int^{\infty}_0 dr_2\  r_2 \int^{2\pi}_0 d\phi_2 \    W_{m_1 n_1}(r_1, \phi + \phi_2) \, W_{m_2 n_2} (r_2, \phi_2)\nonumber \\
  &= \frac{1}{(2\pi)^2} \sum_{m_1, n_1, m_2, n_2=0}^{\infty} \!\!\langle m_1, m_2 |\rho| n_1, n_2 \rangle R_w(m_1,n_1;m_2,n_2) \exp\{i(n_1-m_1)\phi\} \int^{2\pi}_0 d\phi_2\   \exp \left\{i [(n_1+ n_2)-(m_1+m_2)]\phi_2 \right\} \nonumber \\
    &=\frac{1}{2\pi} + \frac{1}{2\pi} \sum_{\substack{ m_1+m_2= n_1+n_2\\ m_1 \neq n_1}} \langle m_1, m_2 |\rho| n_1, n_2 \rangle\    R_w(m_1,n_1;m_2,n_2)\   \exp\{i(n_1 - m_1)\phi\},
    \label{eq:phase_distribution_wigner_QHO}
\end{align}
\end{widetext}
where $ R_w(m_1,n_1;m_2,n_2)$ in Eq.~\eqref{eq:phase_distribution_wigner_QHO} involves an integral over the radial part of the Wigner function and $W_{mn}(x,p)$ is the Wigner function for the operator $|m\rangle \langle n|$  given by \cite{W_mn}
\begin{equation}
W_{mn}(x,p)= \frac{1}{\pi} \int_{-\infty}^{\infty} dy\ \langle x \!-\! y |{m}\rangle\, \langle{n}| x \!+\! y \rangle\,\exp(2ipy).
\end{equation} 
In the second line of Eq.~\eqref{eq:phase_distribution_wigner_QHO}, the integral over the variable $\phi_2$ is non-zero only when the argument of its exponential is zero. This constraint implies that only those density matrix elements $\langle m_1, m_2 |\rho| n_1, n_2 \rangle$ determine the relative phase distribution which satisfy the following relation
\begin{equation}
    m_1+m_2=n_1+n_2\ .
    \label{eqn:QHO_selection_rule}
\end{equation}
All the diagonal density matrix elements, which trivially fulfill this relation, contribute only towards a uniform distribution of $1/2\pi$  and do not carry any information about phase correlations or phase locking. Any non-uniformity in the relative phase distribution $P_w(\phi)$ arises solely from the off-diagonal elements that satisfy Eq.~\eqref{eqn:QHO_selection_rule}. In other words, the existence of coherence between joint Fock states with {\it equal} total excitation is crucial for the emergence of phase locking. These coherence elements therefore serve as the essential quantum resource for establishing phase synchronization between two continuous-variable oscillators. 

\begin{figure}
    \centering
    \includegraphics[width=\linewidth]{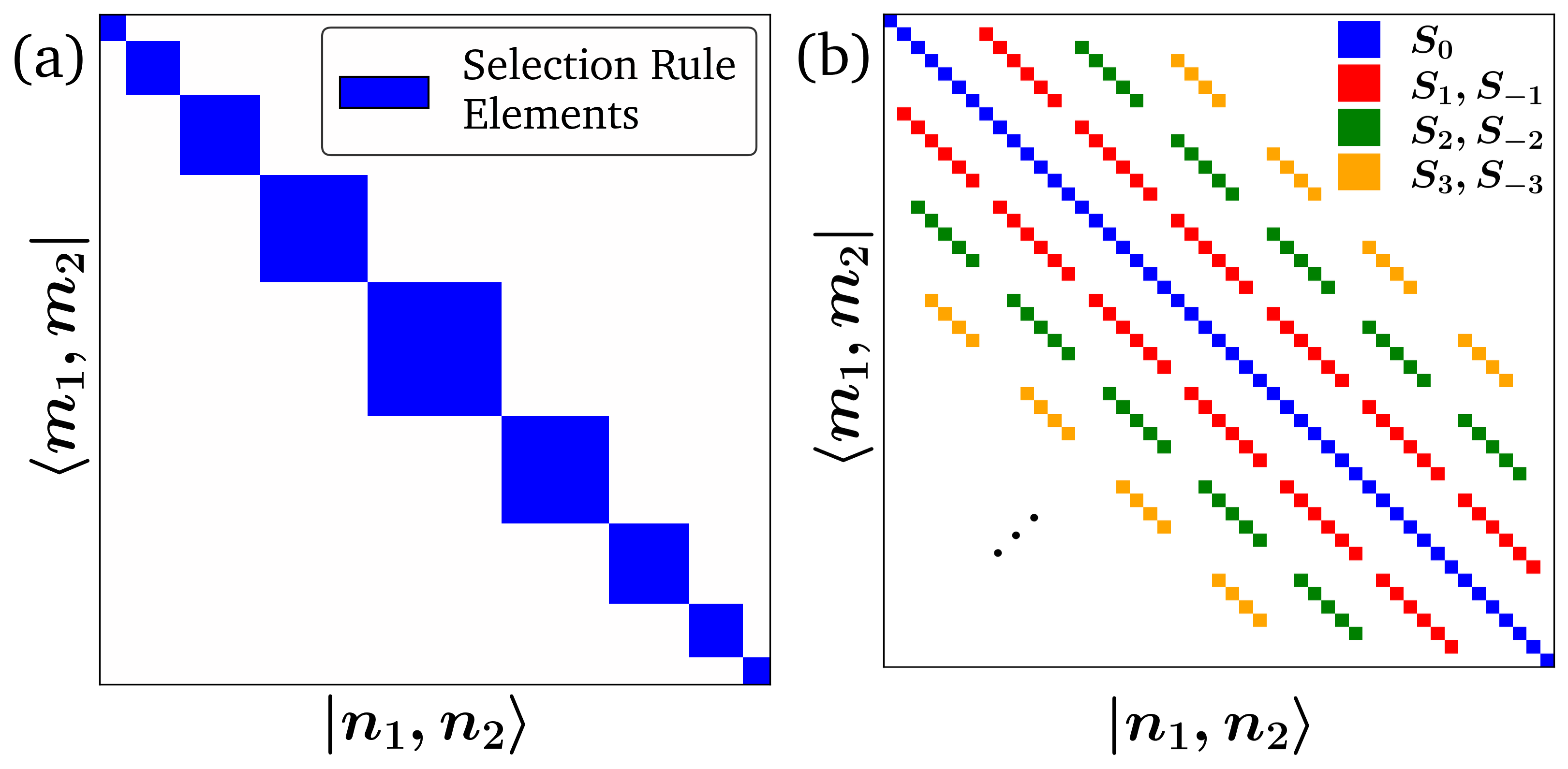}
    \caption{The relative phase distribution between two quantum continuous-variable oscillators is determined by only those density matrix elements $\bra{m_1,m_2}\rho\ket{n_1,n_2}$ which satisfy the excitation relation $(m_1+m_2)=(n_1+n_2)$, as given in Eq.~\eqref{eqn:QHO_selection_rule}. In (a), we reordered the joint Fock basis states such that the states with the same total excitation are grouped together into subspaces. In this representation, the elements following the excitation relation form a block diagonal structure in the density matrix. The coherences within these conserved total excitation subspaces contribute to phase-locking between two CV oscillators. (b) In the joint Fock basis with standard lexicographic ordering, the density matrix elements following the excitation relation are separated into different bands corresponding to subsets $S_k$. Elements of the subset $S_k$ contribute to the $k$-th mode in $P(\phi)$, as given in Eq.~\eqref{eq:P_phi_final}. The plots are generated by truncating the Hilbert space for both oscillators.}
    \label{fig:Selection rule elements}
\end{figure}

For a joint density matrix $\rho$, expressed in a reordered joint Fock basis such that states with same total excitation are grouped together in subspaces, i.e., $\{\ket{0,0};\ket{0,1},\ket{1,0};\ket{0,2},\ket{1,1},\ket{2,0};...\}$, then the matrix elements that follow the condition $(m_1+m_2)=(n_1+n_2)$  form a block diagonal structure, as illustrated in Fig.~\ref{fig:Selection rule elements}(a).
The density matrix elements that satisfy the excitation relation in Eq.~\eqref{eqn:QHO_selection_rule} can be denoted by a set $S$.
This set $S$ can be further partitioned into subsets $\{S_k\}$, where each subset $S_k$, consists of elements for which $m_1-n_1=k$. For instance, all diagonal elements lie in $S_0$, while elements of form $\bra{m_1+1,m_2}\rho\ket{m_1,m_2+1}$, such as $\bra{1,0}\rho\ket{0,1}$, belong to $S_1$. In general, a typical element  in subset $S_k$ has the form $\bra{m_1+k,m_2}\rho\ket{m_1,m_2+k}$. When the density matrix is expressed in the standard lexicographic ordering of the joint Fock basis- i.e., $\{\ket{0,0},\ket{0,1},\ket{0,2},...,\ket{1,0},\ket{1,1},...\}$, these subsets $\{S_k\}$ manifest as distinct bands, as illustrated in Fig.~\ref{fig:Selection rule elements}(b).

One can rewrite the expression for phase distribution $P_w(\phi)$ in Eq.~\eqref{eq:phase_distribution_wigner_QHO} in the form of a harmonic expansion  
\begin{equation}
    P_w(\phi) = \frac{1}{2\pi} + \frac{1}{\pi} \sum_{k}A^w_k \cos(k\phi-\theta^w_k),
    \label{eq:P_phi_final}\end{equation}
where $A^w_k$ denotes the magnitude and $\theta^w_k$ the phase of the complex quantity $C^w_k$, which is defined as the weighted sum over the elements of the subset $S_k$, with the weight factor given by the radial integral $R_w(m_1,n_1;m_2,n_2)$:
\begin{eqnarray}
     C_k^w &=&\sum_{\substack{ m_1+m_2= n_1+n_2\\m_1-n_1=k}}\bra{m_1,m_2}\rho\ket{n_1,n_2} R_w(m_1,n_1;m_2,n_2) \nonumber \\
     & \equiv &A^w_k\,\exp(i\theta^w_k).
     \label{Ckw_main}
\end{eqnarray}
The contribution to the $k$-th harmonic of the relative phase distribution comes from the subset $S_k$. Elements of the subset $S_{-k}$ are simply the complex conjugate of elements of the subset $S_k$, this simplifies the equation of relative phase distribution such that we just require the contributions from the subsets $\{S_k\}$, with $k\ge0$.

For any phase distribution $P(\phi)$, we define a corresponding shifted distribution $S(\phi)$ as, $S(\phi)=P(\phi)-\frac{1}{2\pi}$. The maximum of this shifted distribution, $\max\{S(\phi\}$, serves as a synchronization measure for the system, quantifying the degree of phase localization and hence the strength of synchronization \cite{Spin_correlations,Anharmonic_blockade,micromasers_sync}.

The preceding discussion on phase synchronization was based on the Wigner quasi-probability distribution. However, alternative approaches can also be employed to evaluate the relative phase distribution, such as the Husimi Q-function and the quantum harmonic oscillator (QHO) phase states \cite{SG,PhaseReview,POVM} (see Appendix~\ref{sec: AppendixA} for a detailed derivation). Remarkably, all three methods yield the same excitation condition given in Eq.~\eqref{eqn:QHO_selection_rule} and lead to similar forms for the relative phase distribution $P(\phi)$ in Eq.~\eqref{eq:P_phi_final}. The primary difference among these approaches arises in the complex quantity $C_k$. Husimi Q function has a different radial integral $R_q(m_1,n_1;m_2,n_2)$ which serves as the weight factor for $C_k^q$. Whereas for the case of phase states, $C^p_k$ is simply the sum of all elements in the subset $S_k$, $C_k^p =\sum_{\substack{m_1+m_2= n_1+n_2\\m_1-n_1=k}}\bra{m_1,m_2}\rho\ket{n_1,n_2}$ (see Appendix \ref{sec: AppendixA} for a detailed derivation).

We denote $k_d$ as the mode corresponding to the maximum amplitude, i.e. $A_{k_d}=\max\{A_k\}$. This dominant mode determines the overall structure of $P(\phi)$, resulting in $k_d$ peaks located at  
\begin{equation}
    \phi_P=\frac{(\theta_{k_d}+2n\pi)}{k_d},
\end{equation} 
where $n$ is an integer. The amplitude $A_{k_d}=\max\{A_k \}$, serves as a useful synchronization measure, as it directly relates the density matrix elements to the degree of phase locking.

In the following, we examine various interaction mechanisms proposed in the literature for CV oscillators and analyze how they facilitate phase synchronization through the generation of relevant coherence elements in the joint density matrix. It is important to note that the excitation relation derived here for the CV oscillators, not only offers a fundamental insight into the emergence of phase synchronization, but also provides a practical guideline to engineer mutual interactions and dissipative mechanisms that selectively generate these relevant coherences required for synchronization. 

\begin{figure*}[t]
\centering
\includegraphics[width=\textwidth]{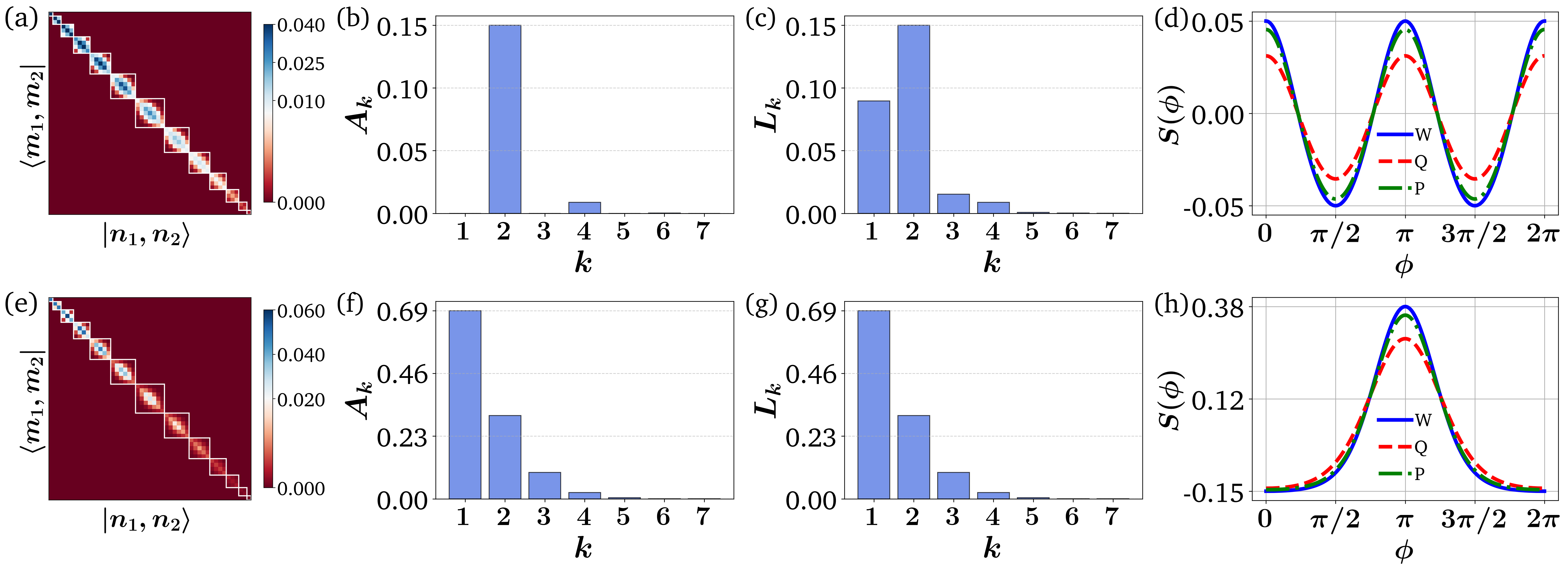}
\caption{Plots for phase synchronization between two VdP oscillators interacting via the coherent coupling of the form $V= g ({a}_1^{\dagger} {a}_2 + {a}_1 {a}_2^{\dagger})$ (a)-(d) and via dissipative coupling of the form $\gamma \mathcal{D}[a_1+a_2]$ (e)-(h). In (a) and (e), we show the color plots of the absolute value of steady-state density matrix elements, for the coherent and the dissipative coupling, respectively. The Fock basis states are reordered, as done in Fig.~\ref{fig:Selection rule elements}(a), and the existence of off-diagonal elements within the conserved total excitation blocks (indicated by white boxes) confirms the generation of coherences following the excitation relation in Eq.~\eqref{eqn:QHO_selection_rule}.  In (b) and (f), we plot $A_k$ (constructed using the quantum oscillator phase states), which is the absolute value for the sum of all elements in the $S_k$ subset. For the coherent interaction (b), the dominant subset is $S_2$. In contrast, for the dissipative case, the dominant subset is $S_1$. In (c) and (g), we plot $L_k$, which is defined as the sum of the absolute values for the elements in the $S_k$ subset. Interestingly, for the coherent case, the $L_k$ value for the $S_1$ subset is non-zero, indicating that the subset has non-zero coherence terms. In contrast, the corresponding $A_k$ is exactly zero, indicating a complete cancellation of coherences within this subset. In (d) and (h), we plot $S(\phi)$ derived from the Wigner Function (W), Husimi Q distribution (Q), and phase states (P). The coherent interaction in (d) leads to a bimodal distribution peaked at $\phi=0$ and $\phi=\pi$. This is because the quantity $A_k$ (and not $L_k$) contributes to phase synchronization and in this case, the dominant contribution comes from $k=2$, which leads to two peaks. In contrast, the dissipative coupling in (h) generates a unimodal relative phase distribution peaked at $\phi=\pi$, as the dominant contribution comes from $A_1$. The parameters are chosen as $\gamma_g^{(1)}=\gamma_g^{(2)}=1$,$\gamma_l^{(1)}=\gamma_l^{(2)}=0.2$, $g=\gamma=1$ and $\omega_1=\omega_2=1$.}
    \label{fig:quantum_vdp}
\end{figure*}

\subsection{Synchronizing Interactions}
{\it Coherent and dissipative coupling between two VdPs:--} In this subsection, we investigate phase synchronization in a system comprising of two VdPs. The dynamics of the uncoupled system is governed by the following Lindblad quantum master equation
\begin{equation}
\dot{\rho} = -i \left[ H_0, \rho \right] + \sum_{i=1,2} \gamma_g^{(i)} \mathcal{D}\left[a_i^\dagger \right] \rho + \gamma_l^{(i)} \mathcal{D}\left[a_i^2 \right] \rho \equiv \mathcal{L}_0 \rho,
\label{eq:QME_two_VdP}
\end{equation}
where $H_0 = \sum_{i=1,2}\omega_ia_i^{\dagger}a_i$ is the free Hamiltonian. The gain and loss rates for each oscillator are given by $\gamma_g^{(i)}$ and $\gamma_l^{(i)}$, respectively. In the absence of any coupling, the steady-state density matrix factorizes as $\rho=\rho_1 \otimes \rho_2$, with each $\rho_i$ being diagonal in the Fock basis. As a result, the relative phase distribution $P(\phi)$ remains flat, reflecting the absence of phase locking between the oscillators. To induce phase synchronization, it is necessary to introduce an interaction that generates coherences within subspaces with conserved total excitation. This can be achieved, for example, through (i) a coherent coupling of the form $V= g ({a}_1^{\dagger} {a}_2 + {a}_1 {a}_2^{\dagger})$ where $g$ is the coherent coupling strength or through (ii) a dissipative coupling via correlated one photon loss, described by the Lindblad term $\gamma \mathcal{D}[{a}_1+{a}_2]$, where $\gamma$ is the strength of the dissipative interaction between the two oscillators. 

Both coherent and dissipative couplings generate coherence elements that satisfy the excitation relation in Eq.~\eqref{eqn:QHO_selection_rule}, as could be seen from Fig.~\ref{fig:quantum_vdp}(a) and (e), respectively. Here we have plotted the absolute values for steady state density matrix elements generated by both kinds of interaction, in the reordered Fock basis such that states with the same total excitation are grouped together into subspaces. Presence of off-diagonal terms in the conserved total excitation blocks confirms that both interactions generate coherences as per the excitation relation in Eq.~\eqref{eqn:QHO_selection_rule}, which leads to phase locking.
However, the two types of interactions give rise to different dominant subsets $S_{k_d}$. In the case of coherent coupling, the dominant subset is $\mathcal{S}_2$, which includes coherences like $\bra{2,0}\rho\ket{0,2}$. This leads to a bimodal phase distribution $P(\phi)$.
In contrast, the dominant subset for the dissipative coupling case is $\mathcal{S}_1$,
resulting in a unimodal distribution, characterized by coherence elements like $\bra{1,0}\rho\ket{0,1}$. We show the plots for relative phase distribution $S(\phi)=P(\phi)-\frac{1}{2\pi}$ in Fig.~\ref{fig:quantum_vdp}(d) and (h) for the coherent and the dissipative case, respectively. We further note that, all three methods -- Wigner distribution (labeled `W'), Husimi distribution (labeled `Q'), and the phase states (labeled `P')-- predict the same peak locations in $P(\phi)$ for this case, with difference arising only in the height of the peaks. 

To further understand the structure of the phase distribution in both cases, we plot the quantity $A_k$ in Fig.~\ref{fig:quantum_vdp}(b) and (f). Note that $A_k$ captures the absolute value of the sum of all the density matrix elements within a given subset $S_k$, and consequently determines the contribution of the $k-$th harmonic mode to $P(\phi)$, as follows from Eq.~\eqref{eq:P_phi_final}. As evident, the dominant subset for the coherent case is $k=2$, while it is $k=1$ for dissipative coupling. Additionally, in Fig.~\ref{fig:quantum_vdp}(c) and (g), we plot the quantity $L_k=\sum_{x\in S_k}|x|$, which measures the absolute strength of the coherence elements in each subset $S_k$. Interestingly, even though the coherence elements with $k=1$ are present in the coherent coupling case (as indicated by non-zero $L_1$), their net contribution $A_1$ vanishes due to destructive interference in the summation [see Eq.~\eqref{eq:P_phi_final}]. Consequently, the subset $S_1$ does not contribute to phase locking in the case of coherent coupling for the chosen set of parameters. However, it is possible to engineer the dissipation rates in such a way that $S_1$ becomes the dominant subset and the destructive interference is suppressed. This, in turn, would lead to a unimodal phase distribution.

It is important to emphasize that while these interactions induce global coherences, the reduced steady states $\tr_1(\rho)$ and $\tr_2(\rho)$, of the individual VdPs remain diagonal in the Fock basis. This implies that these couplings do not induce a definite phase preference for each individual oscillator, but rather they enforce a preference in the relative phase between them. It is worth noting that other forms of dissipative couplings, such as correlated two-photon loss, described by $\mathcal{D}[{a}_1^2+ {a}_2^2]$ or correlated three-photon loss, given by  $\mathcal{D}[{a}_1^3+{a}_2^3]$ can also give rise to nontrivial relative phase distribution. The phase distribution exhibits complex structures, -- two-peak and three-peak distributions, respectively, and the origin of such phase structure can once again be understood from the dominant subset contribution in the joint density matrix. 

\begin{figure}
    \centering
    \includegraphics[width=\linewidth]{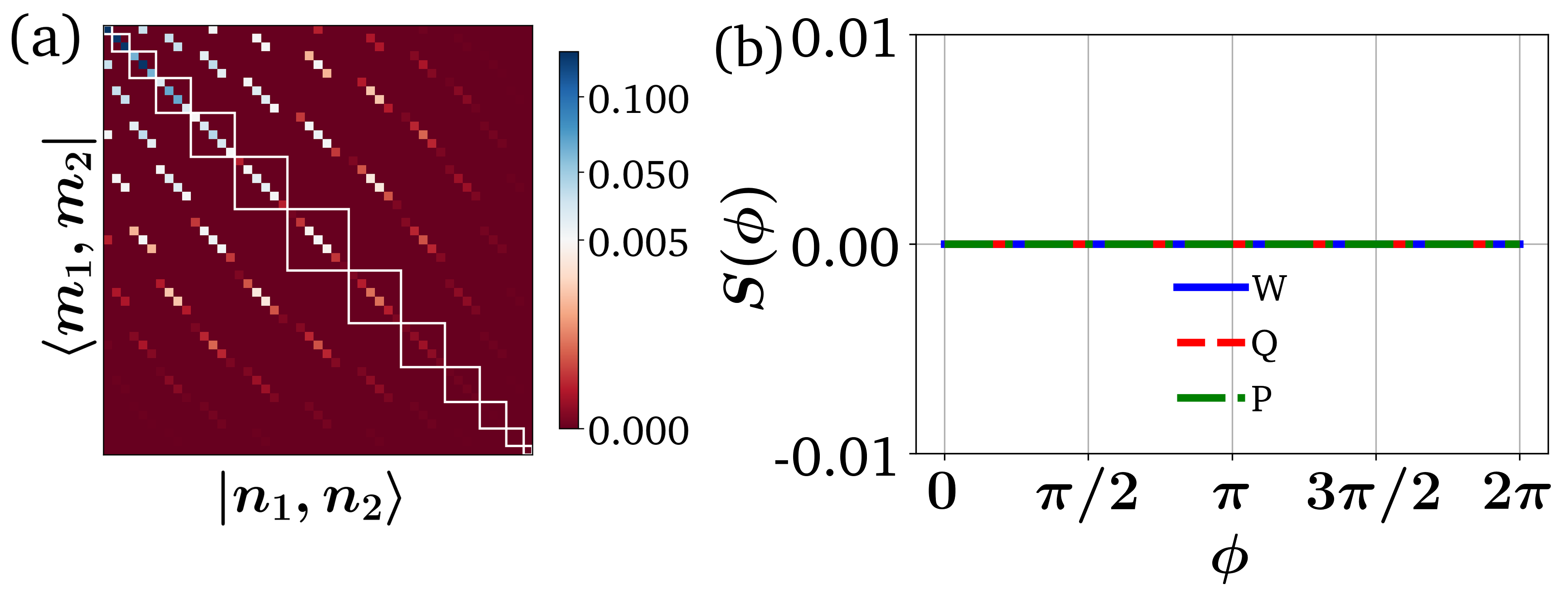}
    \caption{Plots for (a) absolute value of steady-state density matrix elements in the reordered joint Fock basis, as in Fig.~\ref{fig:Selection rule elements}(a), and (b) the relative phase distribution $S(\phi)$ in presence of two-mode squeezing interaction between a pair of VdPs. The form of the interaction is given as $V= ig_s(a_1a_2-a_1^{\dagger}a_2^{\dagger})$. The interaction generates coherences outside the excitation number conserving subspaces, and hence does not lead to phase synchronization. All the three methods --Wigner (W), Husimi (Q), and the phase states (P) -- give a uniform relative phase distribution. The parameters are chosen as $\gamma_g^{(1)}=\gamma_g^{(2)}=\gamma_l^{(1)}=\gamma_l^{(2)}=1$, $g_s=0.5$ and $\omega_1=\omega_2=1$. } 
    \label{fig:Squeezed VdP}
\end{figure}

For a quantum VdP, an interesting special regime is the dissipative limit, which corresponds to taking the ratio $\gamma_l/\gamma_g\to\infty$ in Eq.~\eqref{eq:QME-VdP}. Under this condition, the VdP effectively becomes restricted to its two lowest energy states \cite{1-QHO,Lee-1} and the oscillator could be mapped to a two-level spin ($|0\rangle, |1\rangle$) with dissipators $\mathcal{D}[\sigma_+]$ and $\mathcal{D}[\sigma_-]$, with corresponding rates being $\gamma_+=\gamma$ and $\gamma_-=2\gamma$, respectively \cite{Lee-2}. Here the operators are defined as $\sigma_+=|1\rangle \langle 0|$ and $\sigma_-=|0\rangle \langle 1|$. Previous studies on phase synchronization between two VdP oscillators have shown that, in the dissipative limit, coherent coupling fails to induce phase locking \cite{Lee-1}. This observation can be readily understood following our approach. For the two VdPs in the dissipative limit, the only coherence element following the excitation relation that could exist is $\bra{1,0}\rho\ket{0,1}$ (and its h.c.). Now, the coherent interaction of the form $V=g(\sigma^{(1)}_+\sigma^{(2)}_-+\sigma^{(1)}_-\sigma^{(2)}_+)$, leads to following evolution equation for this coherence element, as obtained following the Lindblad master equation,
\begin{align}
    \frac{d}{dt}\bra{1,0}\rho\ket{0,1}=&\left(-3\gamma+i\Delta\right)\bra{1,0}\rho\ket{0,1}- \nonumber \\
    &ig\Big(\bra{0,1}\rho\ket{0,1}-\bra{1,0}\rho\ket{1,0}\Big),
    \label{eqn:dissipative_ODE}
\end{align}
where $\Delta=\omega_2-\omega_1$ is the detuning between the oscillators. As both the VdPs have similar rates ($\gamma^{(i)}_l/\gamma^{(i)}_g\to\infty$) and are coherently coupled, the master equation remains invariant under the exchange of oscillators $1 \leftrightarrow 2$. This further implies that the following populations are equal, $\bra{0,1}\rho\ket{0,1}=\bra{1,0}\rho\ket{1,0}$. As a result for Eq.~\eqref{eqn:dissipative_ODE}, $\bra{1,0}\rho\ket{0,1}$ decays to zero in the long-time limit, leading to no synchronization. Moreover, for two coherently coupled VdPs with identical rates, the dominant contribution arises from the subset $S_2$, which includes coherence elements such as $\bra{2,0}\rho\ket{0,2}$. The existence of such terms requires access of at least three energy levels - $|0\rangle, |1\rangle, |2\rangle$. However, the dissipative limit confines the system dynamics to just the ground and first excited states. Consequently, the relevant coherence elements cannot form, and phase synchronization does not emerge.

\vspace{0.2cm}
{\it Two-Mode Squeezing interaction between two VdPs :--} We next consider a different form of coherent mutual coupling -- two-mode squeezing interaction-- and examine within our framework, whether it can give rise to phase locking between two VdPs. For a single VdP, it has been shown that squeezing can lead to enhancement of synchronization \cite{Squeezing}. For the case of two VdPs, the Hamiltonian describing the two-mode squeezing interaction is given by, $V= ig_s(a_1a_2-a_1^{\dagger}a_2^{\dagger})$, with $g_s$ being the squeezing strength. The interaction indeed generates coherences between the two oscillators, however, crucially, it does not produce the specific coherence elements that contribute to phase locking. As shown in  Fig.~\ref{fig:Squeezed VdP}(a), the steady state density matrix has all the coherence elements that lie outside the conserved total excitation subspaces. As a result, the relative phase distribution remains uniform, as illustrated in Fig.~\ref{fig:Squeezed VdP}(b). Therefore, in contrast to the single CV oscillator case, the two-mode squeezing does not induce phase synchronization between two CV oscillators.

Perturbation theory provides a useful framework to investigate how different interactions influence a pair of oscillators and whether such interactions can induce synchronization. In this approach, the Liouvillian corresponding to the mutual interaction $\mathcal{L}_I$, is treated as a perturbation to the unperturbed Liouvillian $\mathcal{L}_0$, which describes two non-interacting oscillators. Through perturbative analysis, we find that an interaction can potentially lead to synchronization if the action of $\mathcal{L}_I$ on the steady state of $\mathcal{L}_0$ generates coherence elements that satisfy the excitation relation (see Appendix~\ref{sec:AppendixD} for details).

In what follows, we extend our framework and results to discrete spin systems and to a hybrid configuration involving a spin and an oscillator.

\section{Synchronization of Two discrete Spin systems}
\label{sec:sync-spin}
Discrete variable spin systems can be regarded as self-sustained oscillators, where the notion of phase is naturally described using spin coherent states $\ket{\theta,\phi}$ \cite{1-Spin}. This analogy enables us to perform analogous treatment, as presented in the previous section, to analyze phase synchronization between two self-sustained spin systems. The spin coherent states $\ket{\theta,\phi}$ for arbitrary spin $s$
is constructed by rotating the extremal state $|s, m_s=s\rangle$ -- first counterclockwise by an angle $\theta$ about the $y$-axis, followed by a rotation by an angle $\phi$ about the $z$-axis \cite{LYMAJP_2015,1-Spin}
\begin{equation}
    | \theta, \phi \rangle = \exp(-i \phi {S}_z) \exp(-i \theta {S}_y) |s, m_s=s \rangle.
\end{equation}
The time evolution of the spin coherent state $\ket{\theta,\phi}$ under the free spin Hamiltonian $H_0=\omega_0 {S}_z$ is given by $e^{-i {H}_0 t} \vert \theta, \phi \rangle = \vert \theta, \phi + \omega_0 t \rangle$ which implies that $\phi$ undergoes rotation naturally under the free Hamiltonian and could therefore be treated as a phase variable. 

\subsection{Relative Phase Distribution and The Excitation Relation}
Similar to the case of a CV oscillators, phase synchronization in discrete spin systems can be investigated by analyzing the joint phase-space probability distribution. Specifically, we employ the joint Husimi Q distribution for two arbitrary spins with spin quantum numbers $s_1$ and $s_2$ to extract the information about their relative phase. The joint Q distribution is defined as,
\begin{equation}
Q(\theta_1, \phi_1, \theta_2, \phi_2) =\frac{C_1 C_2}{4 \pi^2} \langle \theta_1, \phi_1; \theta_2, \phi_2 | \rho | \theta_1, \phi_1 ;\theta_2, \phi_2 \rangle, 
\label{eq:example}
\end{equation}
where $C_i = 2 s_i +1$, for $i=1, 2$,  $ | \theta_1, \phi_1 ;\theta_2, \phi_2 \rangle $ is the joint spin coherent state for two spins, and $\rho$ is the joint density matrix. 

To extract information about the relative phase from the Q function, we begin by expanding $\rho$ into the joint basis $|s_1, m_1; s_2, m_2\rangle$, where $\ket{s_i,m_i}$ is the simultaneous eigenbasis for ${S}_{(i)}^2$ and $S_z^{(i)}$. Additionally, we expand $\ket{\theta_i,\phi_i}$ in the $\ket{s_i,m_i}$ basis using the Wigner-D matrix \cite{WignerD, LYMAJP_2015}. We then perform a change of variable from $(\theta_1, \phi_1, \theta_2, \phi_2)$ to $(\theta_1,\theta_2,\phi,\phi_2)$, similar to the case of CV oscillators, where $\phi=(\phi_1-\phi_2)$ is the relative phase. The relative phase distribution $P_{q}(\phi)$ is then obtained by integrating out $\theta_1,\theta_2$, and $\phi_2$ variables (the subscript $q$ indicates that this distribution is derived from the Q function). We write
\begin{equation}
    P_q(\phi) \!=\!\int_0^{\pi}\! d\theta_1 \!\sin \theta_1  \int_0^{\pi}\! d\theta_2 \!\sin \theta_2  \!\int_0^{2\pi} \!d\phi_2 \ Q(\theta_1,\theta_2,\phi,\phi_2), 
\end{equation}
where the $Q$ function can be expressed as (see Appendix \ref{app:spin} for the details of the derivation)
\begin{widetext}
\begin{align}
     Q=  \frac{C_1C_2}{(4\pi)^2} \sum_{m_1,m_2,n_1,n_2} & \bra{s_1,m_1,;s_2,m_2}\rho\ket{s_1,n_1;s_2,n_2} N_{s_1,m_1}(\theta_1) N_{s_2,m_2}(\theta_2)N_{s_1,n_1}(\theta_1)N_{s_2,n_2}(\theta_2) \nonumber \\
     & \exp\{i(m_1 - n_1)\phi\}   \, 
     \exp\{i\left[(m_1+m_2)-(n_1+n_2)\right]\phi_2\}.
     \label{eq:2_spin_Q_func}
 \end{align}
 \end{widetext}
Here $N_{s,m}(\theta)$ is defined as 
\begin{equation}
    N_{s,m}(\theta) = \binom{2s}{s+m} ^{\frac{1}{2}}  \Big(\cos \frac{\theta}{2} \Big)^{s+m} \Big(\sin \frac{\theta}{2}\Big)^{s-m}.
\end{equation}
The summation indices $m_i, n_i$ in Eq.~\eqref{eq:2_spin_Q_func}  takes values $-s_i$, $-s_i+1$, $\cdots,$ $s_i-1$, $s_i$ for $i=1,2$. 
The dependence of $\phi_2$ appears explicitly in the final exponential term in Eq.~\eqref{eq:2_spin_Q_func}. Integration over $\phi_2$ yields a non-zero result only when the argument of its exponential is zero. This means that only those joint density matrix elements $\bra{s_1,m_1;s_2,m_2}\rho\ket{s_1,n_1;s_2,n_2}$ contribute to the relative phase distribution that satisfy the following excitation relation
\begin{equation}
    m_1+m_2=n_1+n_2\ .
    \label{eqn:spin_selection_rule}
\end{equation}
This condition is similar to the one observed for the CV case, although the allowed values of $m_i, n_i$ differ due to the discrete spin nature of the system.
 
A simplified expression for $P_q(\phi)$, derived from the Q function, can be written  as (see Appendix \ref{app:spin} for the details of the derivation),
\begin{widetext}
   \begin{equation}
        P_q(\phi) =  \frac{1}{2\pi} + \frac{C_1 C_2}{8\pi}  \sum_{\substack{ m_1+m_2= n_1+n_2\\ m_1 \neq n_1}} \bra{s_1,m_1;s_2,m_2}\rho\ket{s_1,n_1;s_2,n_2} T_q(m_1,n_1;m_2,n_2)\   \exp\left(i(m_1-n_1)\phi \right),
        \label{eq:prob_phi_spin}
   \end{equation}
\end{widetext}
where the quantity $ T_q(m_1,n_1;m_2,n_2)$ involves integrals over $\theta_1, \theta_2$ variables. The equation for $P_{q}(\phi)$ reveals that if the joint density matrix for two spins is diagonal in the joint basis $\ket{s_1,m_1;s_2,m_2}$, then the relative phase distribution has a uniform value of $1/2\pi$. For phase locking to emerge, the density matrix must contain coherences between states sharing the same total ${S}_z= \sum_{i=1,2} {S}^{i}_z$ eigenvalue, i.e., those satisfying excitation relation in Eq.~\eqref{eqn:spin_selection_rule}. These specific coherence elements act as a key resource for enabling synchronization between two spin systems. The relative phase distribution between two spins in Eq.~\eqref{eq:prob_phi_spin} could be reduced to a superposition of different harmonic modes, similar to Eq.~\eqref{eq:P_phi_final}. The contribution to the $k$-th mode comes from the subset $S_k$, which includes elements for the form  $\bra{s_1,m_1+k;s_2,m_2}\rho\ket{s_1,m_1;s_2,m_2+k}$ (see Appendix \ref{app:spin} for the details).

It is worth emphasizing that, analogous to the case of CV oscillators, one can employ spin phase states to compute the relative phase distribution for spin systems. This alternative approach also yields the same excitation relation in Eq.~\eqref{eqn:spin_selection_rule}, for the joint density matrix elements required for phase synchronization (see Appendix \ref{app:spin} for the details).

\subsection{Synchronizing Interactions}
We now examine the form of mutual interaction between the two spin systems that can lead to phase locking. Our focus is on the synchronization of two spin-1 atoms, each exhibiting a well-defined limit cycle. The dynamics of the uncoupled spin-1 systems is governed by the following Lindblad master equation 
\begin{equation}
    \dot{{\rho}} = -i[H_0, \rho] +\sum_{i=1,2} \gamma_g^{(i)} \mathcal{D}[{S}_+^{(i)} {S}_z^{(i)}] {\rho} + \gamma_l^{(i)} \mathcal{D}[{S}_-^{(i)} {S}_z^{(i)}] {\rho},
    \label{eq:placeholder}
\end{equation}
where $H_0=\sum_{i=1,2}\omega_i {S}_z^{(i)}$ is the bare Hamiltonian of the two spin-1 systems. The steady state for each spin is given by $\ket{0}\langle 0|$, corresponding to an equatorial limit cycle with a free phase $\phi$ \cite{1-Spin}. For compactness, we denote the spin eigenstates as  $\ket{m} \equiv \ket{s,m}$.

\begin{figure}
    \centering
    \includegraphics[width=1\linewidth]{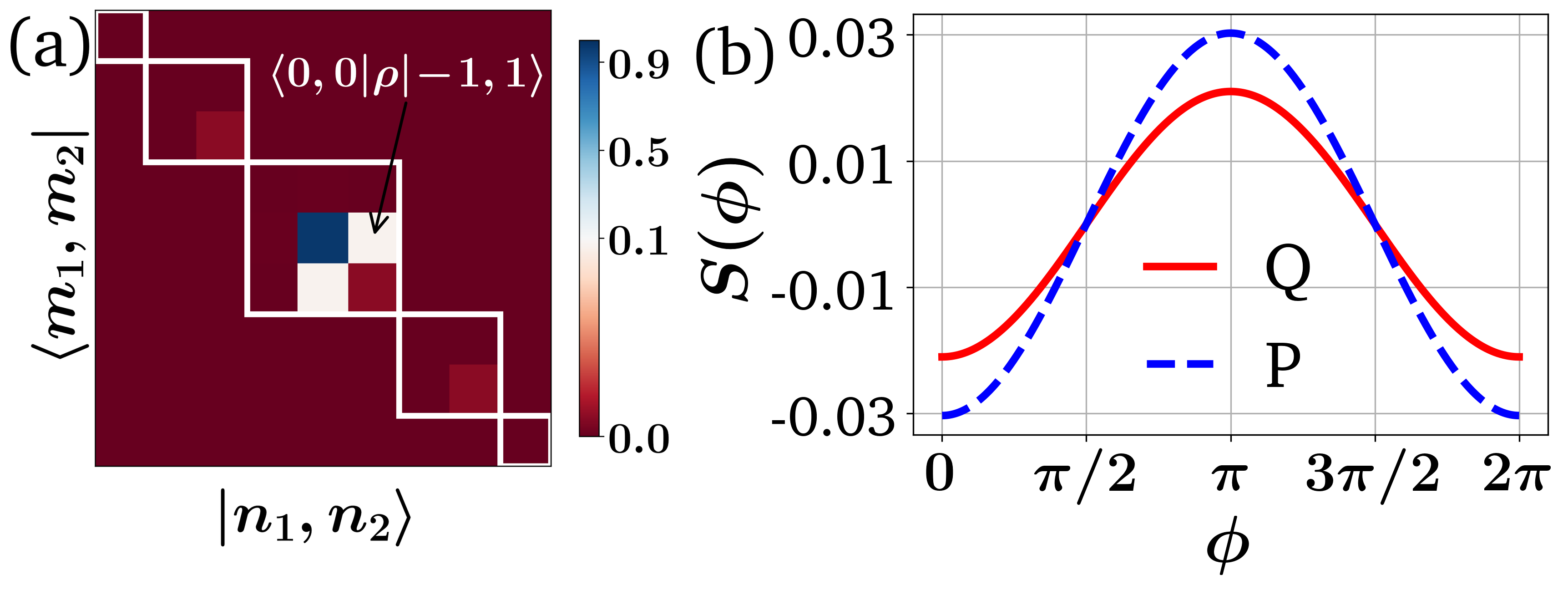}
    \caption{Synchronization between two spin-1 systems with equatorial limit cycle in presence of the coherent interaction $V=i g \left({S}_+^{(1)}{S}_-^{(2)}-{S}_-^{(1)}{S}_+^{(2)}\right)$. (a) Color plot of the absolute values of the steady-state density matrix elements. Here the joint spin states $\{\ket{m_1,m_2}\}$ have been reordered such that states with the same $(m_1+m_2)$ are grouped together. The plot highlights that the dominant contribution arises from the $S_1$ subset, particularly from the coherence element $\bra{0,0}\rho\ket{1,-1}$, which leads to phase locking. (b) The corresponding relative phase distribution exhibits a single peak centered at $\phi=\pi$. The phase distribution is computed using two methods: the Husimi Q-function (labeled as Q) and the spin phase states  (labeled as P), both showing excellent agreement. The parameters used are:  $\gamma_g^{(1)}=\gamma_l^{(2)}=0.01$, $\gamma_l^{(1)}=\gamma_g^{(2)}=1$, $\omega_1=\omega_2=1$ and $g=0.001$.}
    \label{fig:spin_coherent_sync}
\end{figure}

In previous studies, an interaction of the form $V=i g \left({S}_+^{(1)}{S}_-^{(2)}-{S}_-^{(1)}{S}_+^{(2)}\right)$ was considered between two spin-1 systems, and it was shown to induce synchronization \cite{2-Spin}. Within our framework, we identify the mechanism behind this synchronization as the generation of specific coherence elements in the joint density matrix that satisfy the excitation relation of Eq.~\eqref{eqn:spin_selection_rule}. These coherences lie within the subspace conserving the total $S_z$ projection and contribute directly to the phase locking. This is shown in Fig.~\ref{fig:spin_coherent_sync}(a), where the absolute values of the steady state density matrix elements under the coherent interaction are plotted. It is clear that the dominant subset of coherence elements for this interaction corresponds to $S_1$, in particular the coherence matrix elements such as $\bra{0,0}\rho\ket{1,-1}$. As a result, the relative phase distribution becomes unimodal, which is shown in Fig.~\ref{fig:spin_coherent_sync}(b).

\section{Synchronization of Continuous Variable Oscillator and Spin}
\label{sec:sync-hybrid}
So far, we have explored phase synchronization between two continuous-variable oscillators or between two discrete spins. This framework can be suitably extended to a hybrid setup comprising a CV oscillator and a spin \cite{Measures_Sai}, enabling the investigation of phase synchronization in such mixed variable systems as well.

\subsection{Relative Phase Distribution and The Excitation Relation}
As in the previous sections, we adopt the coherent state representations for both the oscillator and the spin. The oscillator is described by the state $|\alpha_o\rangle \equiv\ket{r,\phi_o}$ (with the subscript 'o' denoting the oscillator) and the spin is described by the spin coherent state $|\alpha_s\rangle \equiv |\theta, \phi_s\rangle$ (with 's' representing the spin).
The joint Husimi Q distribution for the hybrid oscillator-spin setup is defined as,
\begin{equation}
    Q(r, \phi_o, \theta, \phi_s) = \frac{2s+1}{4\pi^2} \bra{\alpha_o,\alpha_s} \rho\ket{\alpha_o,\alpha_s},
\end{equation}
where $r$ and $\theta$ encode the populations for the oscillator and the spin, respectively, while $\phi_o$ and $\phi_s$ correspond to their respective phases.

Using the usual Fock basis states for the oscillator $|m_0\rangle$, and spin eigenstates $|s, m_s\rangle \equiv |m_s\rangle$, the Husimi Q function can be expressed as 
\begin{eqnarray}
    Q(r, \phi_o, \theta, \phi_s) &=& \frac{2s+1}{4\pi^2} \sum_{m_o,m_s,n_o,n_s} \bra{m_o,m_s}\rho\ket{n_o,n_s}  \times \nonumber \\
    &&\langle \alpha_o, \alpha_s\ket{m_o,m_s}\bra{n_o,n_s}\alpha_o, \alpha_s\rangle,
    \label{joint-Q-main}
\end{eqnarray}
where $m_o,n_o \in \{0, 1, 2, \cdots \infty\}$ for the oscillator, and $m_s, n_s \in \{-s, -s+1, \cdots, s-1, s\}$ for the spin. We further use the known overlaps between coherence states and the respective basis states, 
\begin{eqnarray}
   \langle n_o\ket{\alpha_o} &=& \frac{\exp\{-r^2/2\}}{\sqrt{n_o!}}\   r^{n_o}\   \exp\{i n_o\phi_o\}, \nonumber \\
   \langle n_s\ket{\alpha_s} &=& N_{n_s}(\theta)\   \exp\{-i n_s\phi_s\}, 
   \label{hybrid}
\end{eqnarray}
where we substitute $\alpha_0=r \, e^{i \phi_o}$, and use the spin coherent state decomposition using the  Wigner-D matrix [see Eq.~\eqref{eq:spin coherent}], to get these results. Similar to the previous cases, we do a variable transform for the Husimi Q function from $Q(r, \phi_o, \theta, \phi_s)$ to $Q(r, \theta, \phi, \phi_s)$, where $\phi=(\phi_o-\phi_s)$ is the relative phase for the hybrid system. To obtain the relative phase distribution we integrate over all variables except $\phi$ (see Appendix \ref{sec:AppendixC} for the details of the derivation)
\begin{widetext}
\begin{align}
    P_q(\phi) &= \frac{2s+1}{8\pi^2} \sum_{m_o,m_s,n_o,n_s} \bra{m_o,m_s}\rho\ket{n_o,n_s}\   \int_0^{\infty} dr\ \frac{1}{\sqrt{m_o!n_o!}} \exp\{-r^2\} \ 2r^{m_o+n_o+1} \   \int_0^{\pi} d\theta\   \sin(\theta)\   N_{m_s}(\theta)N_{n_s}(\theta) \nonumber \\
    & \hspace{3cm} \exp\{i(n_o - m_o)\phi\} \int_0^{2\pi}d\phi_s \,  \exp\{i[ (m_s - n_s)-(m_o - n_o) ]\phi_s\} \nonumber \\
    &=\frac{1}{2\pi} + \frac{2s+1}{2\pi} \sum_{\substack{ m_o - m_s = n_o - n_s\\ m_o \neq n_o}} \bra{m_o,m_s}\rho\ket{n_o,n_s} I_q(m_o,n_o;m_s,n_s) \exp[i(n_o - m_o)\phi].
    \label{phase-hybrid}
\end{align}

\end{widetext}
where the function $I_q(m_o,n_o;m_s,n_s)$ involves integrals over $r$ and $\theta$. As could be seen from Eq.~\eqref{phase-hybrid},  the integral over $\phi_s$ is non-zero only when the argument inside its exponential is zero. This means that only those density matrix elements $\bra{m_o,m_s}\rho\ket{n_o,n_s}$ contribute to the relative phase distribution that satisfy the following excitation relation
\begin{equation}
    m_o-m_s=n_o-n_s\ .
    \label{eqn:hybrid_selection_rule}
\end{equation}

The final form of $P_q(\phi)$ in Eq.~\eqref{phase-hybrid} tells us that a diagonal joint density matrix produces a uniform phase distribution $P(\phi)$. Any non-trivial $\phi$ dependence in the relative phase distribution arises exclusively from the off-diagonal elements satisfying the excitation relation in Eq.~\eqref{eqn:hybrid_selection_rule}. Thus, for the hybrid case, the coherences within subspaces with conserved excitation difference act as the key resource for generating phase locking. The expression for $P_q(\phi)$ could be reduced to a superposition of different harmonic modes, similar to Eq.~\eqref{eq:P_phi_final}.  Here, the contribution to the k-th comes from the subset $S_k$, which includes elements of the form  $\bra{m_o+k,m_s+k}\rho\ket{m_o,m_s}$ (see Appendix \ref{sec:AppendixC} for more details).

One can also use the harmonic oscillator and spin phase states to plot the relative phase distribution for the hybrid case. This approach also gives us the same excitation relation as in Eq.~\eqref{eqn:hybrid_selection_rule} (see Appendix \ref{sec:AppendixC} for more details).

\subsection{Synchronizing Interactions}
We now examine the phase synchronization between an oscillator and a spin-1 system. A widely known form of mutual coupling in such hybrid setup is of Jaynes-Cummings type interaction, given by $V=g({a} {S}_++{a}^{\dagger} {S}_-)$. This interaction conserves the total excitation number and has previously been shown to generate an Arnold tongue structure in the relative entropy of coherence for a hybrid system comprising a VdP and a spin-1 system with equatorial limit-cycle \cite{Measures_Sai}.

\begin{figure}
    \centering
    \includegraphics[width=1\linewidth]{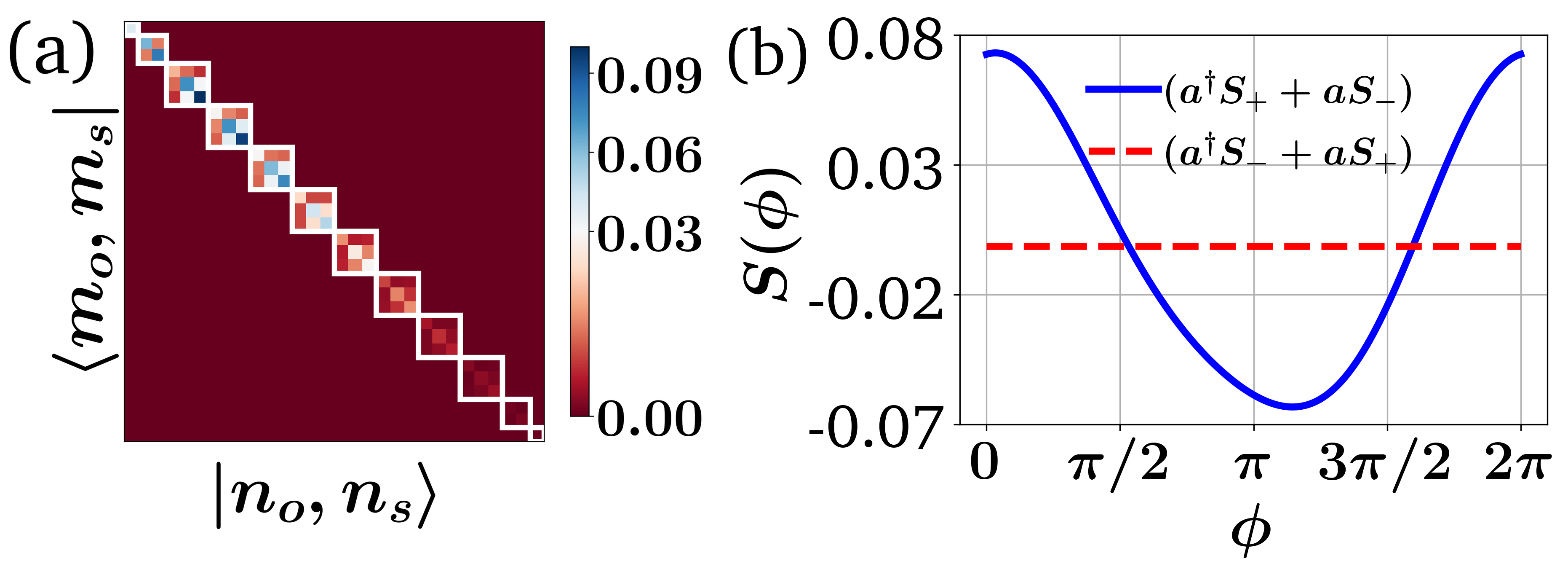}
    \caption{Synchronization between spin-1 system and a VdP interacting via excitation non-conserving coherent interaction  $V=g({a}^{\dagger}{S}_++{a}{S}_-)$. (a) Color plot of the absolute values of steady-state density matrix elements. The basis $\{\ket{m_o,m_s}\}$ is reordered such that states with same $(m_o-m_s)$ are grouped together. This interaction generates coherences as per the excitation relation $(m_o-m_s)=(n_o-n_s)$. (b) Relative phase distribution $S(\phi)$ generated by excitation non-conserving interaction and conserving interaction $V=g({a}{S}_++{a}^{\dagger}{S}_-)$. As can be seen, only the former form of interaction generates phase synchronization. The parameters are chosen as $\gamma^o_l=0.2,\gamma^o_g=1$ for the VdP and  $\gamma^s_l=0.01,\gamma^s_g=1$, for the spin-1 with equatorial limit-cycle, $\omega_o=\omega_s=1$ and $g=1$.}  
    \label{fig:Hybrid_system}
\end{figure}

Note that for observing phase synchronization, the excitation relation for the hybrid system, Eq.~\eqref{eqn:hybrid_selection_rule}, is quite different from the excitation relations for similar systems, Eq.~\eqref{eqn:QHO_selection_rule} and Eq.~\eqref{eqn:spin_selection_rule}. 
As for the hybrid setup, the density matrix elements that exist within the subspaces with conserved excitation difference are responsible for synchronization, that's why the total excitation conserving interaction $V=g({a}{S}_++{a}^{\dagger}{S}_-)$ fails to generate phase synchronization. On the other hand, interaction of the form $V=g ({a}^{\dagger}{S}_++{a}{S}_-)$ that does not conserve total excitation but instead allows for simultaneous excitation and de-excitation of both subsystems generates the required coherences as per the excitation relation in Eq.~\eqref{eqn:hybrid_selection_rule}. This interaction therefore enables synchronization between the spin and the oscillator. We define the joint basis for hybrid case as $\{\ket{m_o;s,m_s}\}\equiv\{\ket{m_o,m_s}\}$ and reorder the basis such that states with the same excitation difference $(m_o-m_s)$ are grouped together into subspaces. In Fig.~\ref{fig:Hybrid_system}(a), we present a color plot for the absolute values of the steady state density matrix under the excitation non-conserving interaction, in the reordered basis.  Coherences appear within the subspaces of fixed excitation difference, confirming the condition for synchronization. This leads to non-uniform relative phase distribution, as shown in Fig.~\ref{fig:Hybrid_system}(b). For comparison, we also show the relative phase distribution $S(\phi)$, resulting from coherent excitation conserving interaction. As this interaction does not generate the relevant coherence elements, the phase distribution remains flat, indicating the absence of synchronization.

\section{Coherence and Correlations}
\label{sec:coh_corr}
 As shown in the previous sections, phase synchronization between two quantum self-sustained oscillating systems requires the existence of specific coherence elements in the joint density matrix. This statement leads to certain important conclusions related to characterizing bipartite synchronization.

\subsection{Coherence and Synchronization}
Recent studies have demonstrated that, for a single oscillator driven by an external source, coherences between all energy eigenstates serve as a resource for synchronization \cite{Resources1SSO,Half_integer}. In contrast, as discussed in the previous sections, the mechanism of synchronization between two quantum SSOs exhibits a markedly different behaviour. Specifically, only the coherence elements within the subspaces that conserve total excitation or excitation difference, contribute to phase locking. 
This indicates that the presence or absence of {\it all other} coherence components has no bearing on the relative phase distribution. More generally, it is therefore possible for the two oscillators to develop substantial mutual coherence without exhibiting any phase locking. 

A generalized measure of bipartite synchronization was recently proposed in Ref.~\cite{Measures_Sai}, which is based on the distance between a quantum state $\rho$ and its nearest reference limit-cycle state of the form $\sigma=\sum_i q_i \, \sigma_i^{(1)} \otimes \sigma_i^{(2)}$, where $\sigma_i^{(\alpha)}$ denotes  the individual limit-cycle state of oscillator $\alpha$ ($\alpha=1,2 $) and the coefficients satisfy $0 < q_i <1$. The synchronization measure is defined via the quantum relative entropy  $\Omega_R(\rho)=\min_{\sigma} S(\rho||\sigma)$, where $S(\rho||\sigma)=\mathrm{Tr}(\rho\ln\rho-\rho\ln\sigma)$. Interestingly, for systems with diagonal limit-cycle states (i.e., $\sigma^{1,2}_i$ are diagonal), as is the case for two uncoupled VdP oscillators or two uncoupled spin-1 systems, the quantity $\Omega_R(\rho)$ reduces to the relative entropy of coherence, 
$\Omega_R(\rho)= S_{\rm coh}(\rho) = S(\rho_{\rm dia})-S(\rho)$,
where $S(\rho)=-\mathrm{Tr}(\rho\ln \rho)$ is the von Neumann entropy and $\rho_{\rm dia}$ is the diagonal density matrix obtained by setting all the off-diagonal terms to zero. Therefore, $S_{\rm coh}(\rho)$ captures {\it all} coherences that could exist between two quantum systems, including the ones that do not contribute to phase synchronization. 

\begin{figure}[h]
    \centering
    \includegraphics[width=1\linewidth]{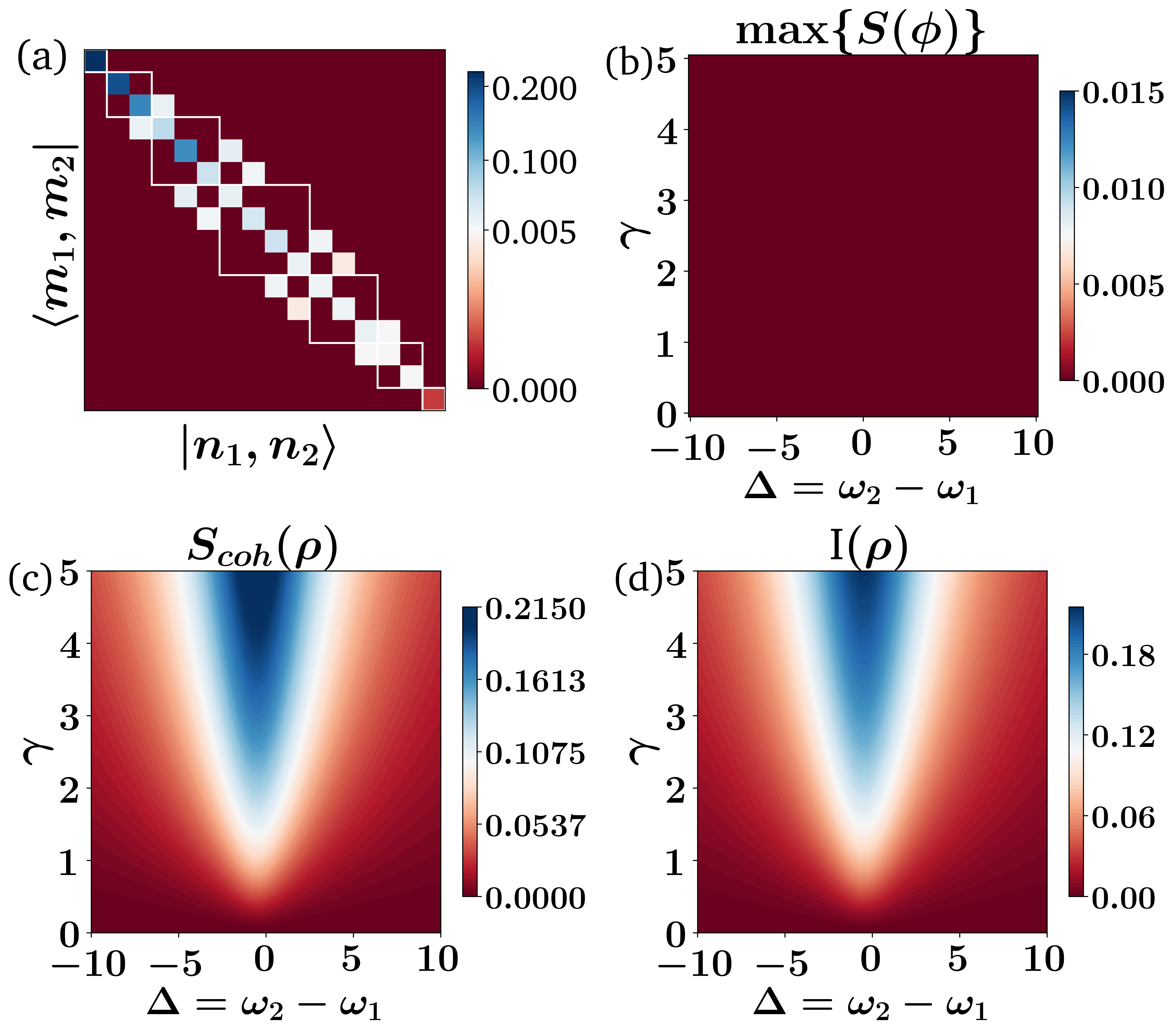}
    \caption{Plot for (a) absolute values of the steady state density matrix elements for a system of two VdPs coupled via a correlated one and two-photon loss, which is modeled by a dissipator of the form $\gamma \mathcal{D}[a_1+a_2^2]$. This coupling leads to the generation of coherences outside the total excitation conserving subspaces. As a result, a flat relative phase distribution is observed, even after having the presence of coherences. This is shown by a two-dimensional plot of the synchronization measure $\max\{S(\phi)\}$, as a function of $\gamma$ and $\Delta=\omega_2-\omega_1$, under the same interaction. As can be seen,  the oscillators remain unsynchronized throughout the whole parameter regime. (c) Plot for the relative entropy of coherence, $S_{\rm coh}(\rho)$, which shows an Arnold tongue behaviour even in the absence of phase synchronization. (d) Plot for mutual information $I(\rho)=S(\rho_1)+S(\rho_2)-S(\rho)$ which also shows Arnold tongue structure in the absence of synchronization. The parameters are chosen as $\gamma_g^{(1)}=\gamma_g^{(2)}=\gamma_l^{(1)}=\gamma_l^{(2)}=1$ and for (a) $\omega_1=\omega_2=1$, $\gamma=1$.}
    \label{fig:Non-sync Int.}
\end{figure}

Let us consider a scenario where two independent VdP oscillators with diagonal steady-states are coupled through correlated one and two-photon loss processes. This interaction can be modeled using a Lindblad dissipator of the form $\gamma \mathcal{D}[{a}_1+{a}_2^2]$. Such a coupling generates coherences in the joint density matrix that exist outside the conserved total occupation subspaces. This behaviour is clearly demonstrated in Fig.~\ref{fig:Non-sync Int.}(a). As a consequence, despite the presence of coherences, this form of coupling does not lead to phase synchronization, as evidenced in Fig.~\ref{fig:Non-sync Int.}(b) where we plot ${\rm max}\{S(\phi)\}$. Nevertheless, the measure $\Omega_R(\rho)=S_{\rm coh}(\rho)$ still exhibits an Arnold tongue structure, as depicted in Fig.~\ref{fig:Non-sync Int.}(c), due to its sensitivity to all forms of coherence -- whether or not they contribute to synchronization. This observation highlights a key limitation of $\Omega_R(\rho)$ as it captures overall coherences rather than the synchronization relevant coherences. To make $\Omega_R(\rho)$ a more reliable measure for phase synchronization, the set of reference states $\sigma$ should be extended to include all density matrices that give rise to a uniform relative phase distribution $P(\phi)$. Importantly, such reference states, in principle, could be entangled and need not be of the separable product form, originally considered.

\begin{figure}[h]
    \centering
    \includegraphics[width=1\linewidth]{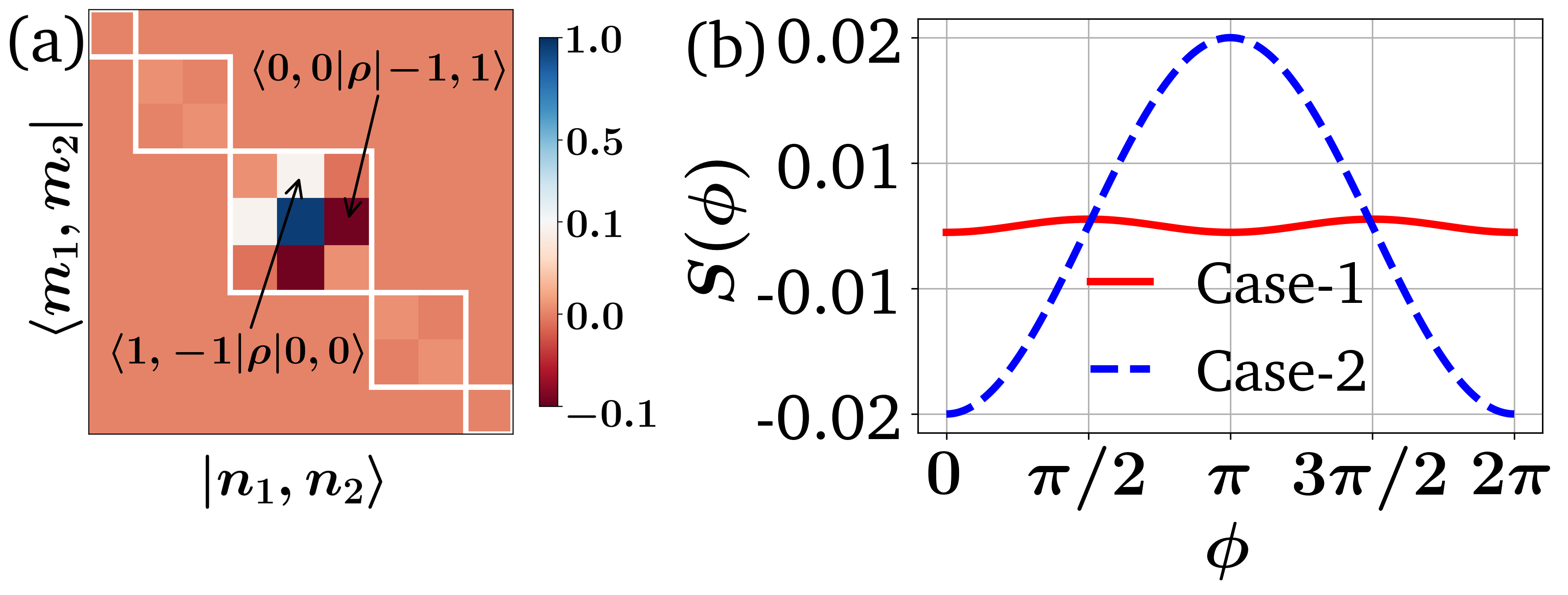}
    \caption{(a) Color plot for the steady state density matrix elements for two coherently coupled spin-1 systems with equatorial limit cycle with $\gamma_g^{(1)}=\gamma_l^{(2)}= \gamma_l^{(1)}=\gamma_g^{(2)}=1$, $\omega_1=\omega_2=1$ and $g=0.1$. The basis is reordered similar to Fig.~\ref{fig:spin_coherent_sync}(a).  The coherences within the subset $S_1$ have opposite signs and cancel each other out. As a consequence of such destructive interference, the relative phase distribution has no contribution from $\cos\phi$, $S_2$ becomes the dominant subset, which leads to a bimodal phase distribution (red solid line), as observed in (b). Once the loss and gain rates are unequal, a unimodal distribution emerges, as shown in Fig.~\ref{fig:spin_coherent_sync}(b) and is also shown here for comparison.}
    \label{fig:Interefrence}
\end{figure}

It is also important to highlight that the mere presence of coherence elements in the joint density matrix, those satisfying the excitation relations, does {\it not} always guarantee the emergence of phase synchronization. This can be understood as follows. The contribution to the $k$-th mode of the relative phase distribution $P(\phi)$ depends on the (weighted) sum of all the elements within that subset $S_k$. Therefore, in certain situations, destructive interference between the coherence terms within a given subset can suppress their net contribution to phase synchronization. This feature was previously noted in Ref.~\cite{2-Spin} for two spin-1 systems coupled via the coherent interaction $V=i g \left({S}_+^{(1)} {S}_-^{(2)}- {S}_-^{(1)} {S}_+^{(2)}\right)$ with all gain and loss rates being equal i.e., $\gamma_g^{(1)}=\gamma_l^{(2)}= \gamma_l^{(1)}=\gamma_g^{(2)}$. The underlying reason for the absence of synchronization in this scenario is now clear from the structure of the steady-state density matrix. Coherence elements within the subset $S_1$, $\bra{0,0}\rho\ket{-1,1}$ and $\bra{1,-1}\rho\ket{0,0}$, appear with equal magnitude but with opposite signs, leading to a destructive interference.  This is shown in Fig.~\ref{fig:Interefrence}(a). As a result, $P(\phi)$ shows no contributions from the first harmonic $\cos(\phi)$. The dominant contributions then arises from the elements belonging to the second subset i.e., $S_2$, leading to a bimodal phase distribution, as can be seen in  Fig.~\ref{fig:Interefrence}(b) (red solid line). Note that such destructive interference effect disappears once the loss and gain rates become asymmetric and a unimodal phase distribution emerges, as seen earlier in Fig.~\ref{fig:spin_coherent_sync}(b) and is also shown in Fig.~\ref{fig:Interefrence}(b) (blue dashed line) for the purpose of comparison.

\subsection{Correlations and Synchronization}
\label{corr_sync}
Synchronization between two quantum systems is accompanied by the development of correlations between them. Because of this mutual information was proposed as a measure for synchronization \cite{Mutual_info}. Given a joint density matrix $\rho$, mutual information between a bipartite system is defined as $I(\rho)=S(\rho_1)+S(\rho_2)-S(\rho)$, where recall that $S$ stands for von Neumann entropy and $\rho_{i}$ is the reduced density matrix of individual subsystem. This quantity captures all forms of possible correlations between two systems, but correlations in general don't always mean synchronization. For example in Fig \ref{fig:Non-sync Int.}(d), the system of two VdPs coupled via $\gamma \mathcal{D}[{a}_1+{a}_2^2]$ shows an Arnold tongue in mutual information but generates no phase synchronization in the entire parameter regime.

A special form of quantum correlation is quantum entanglement. The relation between synchronization and entanglement has been studied previously in various setups \cite{2-QHO,Lee-2}. The fact that synchronization depends on specific density matrix elements, clearly indicates that not all entangled states would exhibit phase locking. As a simple example, consider a bipartite setup consisting of two spin-1 systems prepared in the state  $\ket{\psi}=\big(\ket{1,1}+\ket{0,0}+\ket{-1,-1}\big)/\sqrt{3}$ which is maximally entangled, but such a state would not give rise to any phase locking as the corresponding density matrix $\rho=\ket{\psi}\bra{\psi}$ shows no coherences as per the excitation relation in Eq.~\eqref{eqn:spin_selection_rule}. Whereas the density matrix for state  $\ket{\psi}=\big(\ket{1,-1}+\ket{0,0}+\ket{-1,1}\big)/\sqrt{3}$ has the corresponding total excitation conserving elements (total $S_z$ quantum number is conserved) and can exhibit non-uniform phase distribution $P(\phi)$. 

\begin{figure}[h]
    \centering
    \includegraphics[width=1\linewidth]{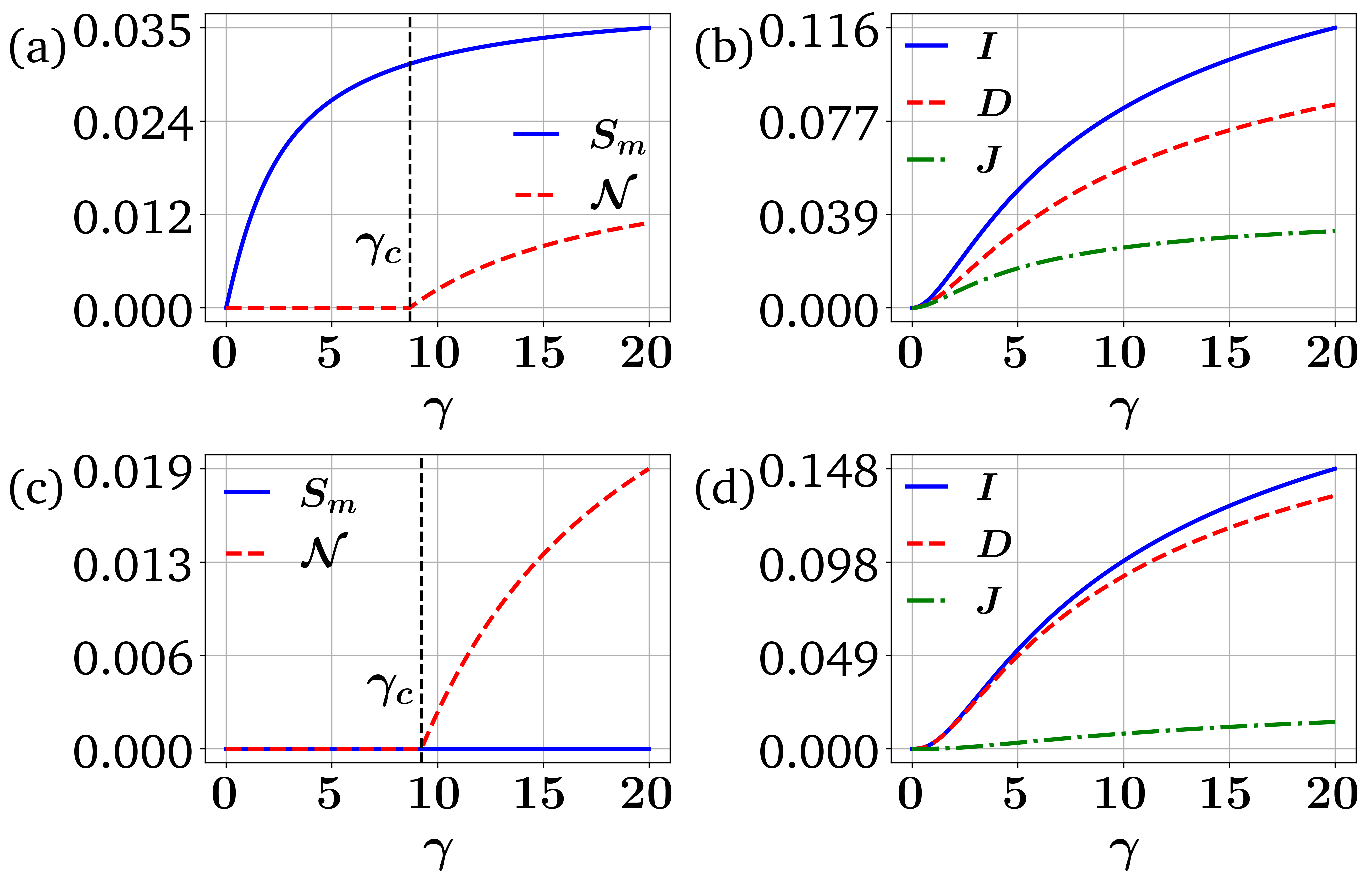}
    \caption{Plots of various measures for two VdPs in the dissipative limit with $\omega_1=\omega_2=1$. In this limit, each oscillator is confined in the lowest two levels, making negativity $\mathcal{N}$ a faithful measure of entanglement. (a) Plot for negativity $\mathcal{N}$ and $S_m=\max\{S(\phi)\}$ in presence of dissipative interaction of the form $\gamma \mathcal{D}[{a}_1+{a}_2]$ as a function of  coupling strength $\gamma$. The synchronization measure $S_m$ is finite in a region where $\mathcal{N}=0$, suggests the presence of synchronization even if the oscillators are unentangled. (c) Plot for the same but in presence of a coupling of the form $ \gamma  \mathcal{D}[{a}^{\dagger}_1+{a}_2]$ and we observe the presence of entanglement but zero synchronization which shows that it is possible for two entangled oscillators not to be synchronized at all. In (b) and (d) we plot the mutual information $(I)$, classical information $(J)$ and quantum discord $(D)$ between two VdPs for interactions $\mathcal{D}[{a}_1+{a}_2]$ and $\mathcal{D}[{a}^{\dagger}_1+{a}_2]$, respectively. Correlations build up between the oscillators for both cases, with classical and quantum contributions.} 
    \label{fig:Entanglement}
\end{figure}

One can make this point more concrete by considering two VdP oscillators in the dissipative limit \cite{Lee-2} i.e., taking $\gamma^i_l/\gamma^i_g\to\infty$ for each oscillator in Eq.~\eqref{eq:QME_two_VdP}.  This leads to the occupation of the lowest two levels for each oscillator. For such two two-level systems, negativity is a faithful measure of entanglement which is defined as $\mathcal{N} = \left( \left\| {\rho}^{T_1} \right\|_1 - 1 \right) / 2$, where $\left\| {\rho}^{T_1} \right\|_{1}$ is the trace norm of partially transposed density matrix over the first VdP.
In presence of a correlated one photon loss interaction of the form $\gamma \mathcal{D} [{a}_1+{a}_2]$, we observe that upon increasing the coupling strength $\gamma$, the synchronization measure $S_m={\rm max}\{S(\phi)\}$ starts to grow instantly, as shown in Fig.~\ref{fig:Entanglement}(a). However, negativity ${\cal N}$ remains zero initially and starts to increase abruptly after a critical strength $\gamma_c$. This behaviour clearly demonstrates that synchronization can exist without entanglement. 

One can further calculate quantum discord which captures all kinds of quantum correlations beyond entanglement. Discord is given as the difference between total correlations $[I(\rho)]$ and the classical correlations $[J(\rho)]$, which is the maximum amount of information that could be obtained about a system by making measurements on the other system. This provides us with an expression for the discord as $D(\rho)=I(\rho)-J(\rho)$ \cite{X_states,Q.Correlations,Q.Discord}, where the measurements are made on the second system. The synchronized states of two VdPs in the dissipative limit come under the category of X states, i.e., the states with non-zero elements are only along the diagonal and anti-diagonal. The calculation of discord for such X states is straightforward \cite{X_states}. In Fig.~\ref{fig:Entanglement}(b) we show that, as $\gamma$ increases, the mutual information ($I$) increases between the oscillators. While mutual information has both, classical and quantum contributions, major weightage is of quantum correlations. Interestingly, this shows that even when two synchronized oscillators are not entangled, there could exist other forms of quantum correlations between them.

Recall that, we showed that the presence of synchronization does not necessarily imply entanglement. The reverse statement is also valid. It is possible that two systems having finite entanglement but no synchronization.  In presence of coupling of the form $\gamma \mathcal{D}[{a}_1^{\dagger}+{a}_2]$ between the two VdPs in the dissipative limit, coherences survive in the steady state which lead to entanglement but these are not the elements which follow the excitation relation. As observed in Fig.~\ref{fig:Entanglement}(c), after the critical coupling $\gamma_c$, the oscillators become entangled but synchronization measure remains zero throughout. This implies that the existence of entanglement doesn't necessarily imply synchronization. For the same coupling, we once again compute the quantum and classical correlations and observe their existence. 

It is clear from this analysis that the presence of measures like relative entropy of coherence, mutual information, or quantum entanglement may not always be linked with synchronization. A faithful measure of synchronization for these kinds of self-sustained oscillators could be the quantity $A_{k_d}$, which can be accessed from the density matrix elements.

\section{Summary}
\label{sec:summary}
We investigate the origin of phase synchronization between two quantum self-sustained oscillators, considering three different configurations: continuous-variable systems, discrete spin systems, and a hybrid setup involving a spin and an oscillator. By analyzing the quasi-probability distribution functions, we identify a simple yet general condition to distinguish the elements of the joint density matrix that serve as a key resource for the emergence of phase synchronization. The condition reflects that the coherences within the subspaces that conserve the total excitation (excitation difference) for the continuous or discrete (hybrid) setup are the sole contributors to the phase synchronization. This insight is valuable for two main reasons: (i) it reveals what forms of interactions could generate the relevant density matrix elements leading to phase synchronization, and (ii) it uncovers a connection between phase synchronization and various information-theoretic measures. We validate our theory using several well-known interaction models-- both coherent and dissipative-- that induce synchronization between quantum self-sustained oscillators, and we also identify interaction forms that fail to generate synchronization. Additionally, we demonstrate that certain information-theoretic measures can sometimes be misleading when used as indicators for synchronization, and we provide a faithful measure of synchronization, $A_{k_d}$.

While the present work focuses on relative phase synchronization characterized by the relative phase $\phi=\phi_1-\phi_2$, our framework can be naturally extended to study more general $n:m$ phase locking, where the relevant phase variable is defined as $\phi=(n\phi_1-m\phi_2)$ \cite{Pikovsky,SimpleToComplex}. Such generalized synchronization may arise from a different set of coherence elements than those responsible for the $1:1$ phase locking, discussed in this paper. As a result, the resources underpinning synchronization could differ significantly, potentially necessitating novel forms of interaction to generate the appropriate coherence elements. An illustrative example is provided in Ref.~\cite{Quadratically}, where $2:1$ phase locking between two coupled VdPs is achieved via an interaction of the form $(a_1^{\dagger2}a_2+a_1^2a_2^{\dagger})$.

It will also be interesting to generalize this framework to more complex, extended systems with larger degrees of freedom, and to investigate the generation of various forms of coherences arising from combinations of different interactions and drives.

\section*{Acknowledgments}
BKA acknowledges CRG Grant No. CRG/2023/003377 from Anusandhan National Research Foundation (ANRF), Government of India.



\onecolumngrid
\appendix
\setcounter{equation}{0}
\setcounter{figure}{0}

\section{Phase distribution of continuous variable oscillator}
\label{sec: AppendixA}
In this appendix, we give a detailed derivation for the phase distributions for continuous-variable oscillators, and arrive at the same excitation relation for the phase synchronization, starting from three different approaches: (i) Wigner function, (ii) Husimi Q function, and (iii) harmonic oscillator phase state approach.

\subsection{Wigner Function}
{\it Single oscillator case.--}  In this subsection, we first discuss the Wigner function approach for the single oscillator case.  Wigner function for the state $\rho$ of a continuous variable oscillator is defined as
\begin{equation}
    W(x, p) = \frac{1}{\pi} \int_{-\infty}^{\infty} dy\ \langle x - y | \rho | x + y \rangle\ \exp(2ipy). 
    \label{eq:wigner def 1 QHO}
\end{equation}
By expanding the density matrix in the Fock basis of harmonic oscillator, we obtain
\begin{align}
    W(x, p) &= \sum_{m,n=0}^{\infty} \bra{m}\rho\ket{n}\ \left[ \frac{1}{\pi} \int_{-\infty}^{\infty} dy\ \langle x - y \ket{m}\bra{n} x + y \rangle\ \exp(2ipy) \right], \nonumber \\
    &= \sum_{m,n=0}^{\infty} \bra{m}\rho\ket{n}\ W_{mn}(x,p),
    \label{eq:Wig fock-1}
\end{align}
where an explicit expression for $W_{mn}(x,p)$ is  given by \cite{W_mn}
\begin{equation}
     W_{mn}(x,p) = \sqrt{\frac{n!}{m!}} \, e^{i(n-m) \arctan(p/x)} \, \frac{(-1)^n}{\pi} \left( 2(x^2 + p^2) \right)^{(m - n) / 2} L_{n}^{m-n}\left( 2(x^2 + p^2)\right) e^{-(x^2 + p^2)}.
     \label{eq:W-mn}
\end{equation}
Here $L_n^{m-n}$ is the generalized Laguerre polynomial. To find the angular distribution, we perform a variable transform to polar coordinates $(r,\phi)$ and integrate over the radial coordinate. We define
\begin{equation}
    r = \sqrt{x^2+p^2}\  ;  \  \phi= \arctan\left(\frac{p}{x}\right)\hspace{-0.4cm}\mod 2\pi.
    \label{eq: def r, phi}
\end{equation}
The marginal probability distribution for angular coordinate $P_w(\phi)$ is then given as
\begin{align}
     P_w(\phi) &=\int_0^{\infty} dr \ r \ W(r,\phi) \nonumber \\
     &=\sum_{m,n=0}^{\infty} \bra{m}\rho\ket{n} \int_0^{\infty}dr\ r\ W_{mn}(r,\phi).
     \label{pwphi}
\end{align}
The subscript $w$ in $P_w(\phi)$ refers to the fact that the distribution is computed using Wigner distribution. Following Eq.~\eqref{eq:W-mn}, in terms of the polar coordinate variable, $W_{mn}(r,\phi)$ could be written as
\begin{equation}
     W_{mn}(r,\phi) =\frac{1}{2\pi} e^{i(n-m) \phi}\times 2(-1)^n \sqrt{\frac{n!}{m!}} \left( 2r^2 \right)^{(m - n) / 2} L_{n}^{m-n}\left( 2r^2\right) \ e^{-r^2}.
     \label{eq:W-mn_radial}
\end{equation}
Using Eq.~\eqref{eq:W-mn_radial} and Eq.~\eqref{pwphi}, $P_w(\phi)$ can then be expressed as
\begin{align}
     P_w(\phi)= \frac{1}{2\pi} \sum_{m,n=0}^{\infty} \langle m | \rho | n \rangle \  r_w(m,n)\  \exp\{i(n-m)\phi \},
     \label{eq:1 HO P(phi) coh}
\end{align}
where $r_w(m,n)$ is the integral of the radial part for the Wigner distribution that follows from Eq.~\eqref{pwphi} and Eq.~\eqref{eq:W-mn_radial}. Its explicit form is given by
\begin{equation}
    r_w(m,n)= 2(-1)^n \sqrt{\frac{n!}{m!}}  \int^{\infty}_0 dr\  r  \left( \sqrt{2}r \right)^{(m - n)} L_{n}^{m-n}( 2r^2) \,  e^{-r^2}.
    \label{rwmn}
\end{equation}
For a Fock state $\ket{m}$, its Wigner distribution $W_{mm}(r,\phi)$ is normalized over the phase space. This gives us the property $r_w(m,m)=1$. Moreover, because of the condition $W_{mn}=W_{nm}^*$, we have $r_w(m,n)=r_w(n,m)$. For later purposes, we next define a complex quantity $C_k^w$ as
\begin{equation}
     C_k^w = \sum_{m-n=k}\bra{m}\rho\ket{n}r_w(m,n),
     \label{eq:ckw}
\end{equation}
which could be decomposed into its radial and angular parts as $C_k^w=A_k^w\exp(i\theta_k^w)$. Using these definitions and properties of $r_w(m,n)$, the equation for $P_w(\phi)$ in Eq.~\eqref{eq:1 HO P(phi) coh} can be simplified as
\begin{align}
    P_w(\phi)&=\frac{1}{2\pi} \sum_{m,n=0}^{\infty} \langle m | \rho | n \rangle \  r_w(m,n)\  \exp\{i(n-m)\phi \}\nonumber \\
    &=\frac{1}{2\pi} + \frac{1}{2\pi} \sum_{m \neq n} \langle m | \rho | n \rangle  r_w(m,n)\  \exp\{i(n-m)\phi \}\nonumber \\
    &=\frac{1}{2\pi} + \frac{1}{2\pi} \sum_{m > n} \langle m | \rho | n \rangle  r_w(m,n)\  \exp\{i(n-m)\phi \} + \langle n | \rho | m \rangle  r_w(n,m)\  \exp\{i(m-n)\phi \}\nonumber \\
    &=\frac{1}{2\pi} + \frac{1}{2\pi}\sum_{k=1}^{\infty} \left[ \exp(-ik\phi )\sum_{m-n=k}\bra{m}\rho\ket{n}r_w(m,n)\right] +h.c.
    \nonumber \\
    &= \frac{1}{2\pi} + \frac{1}{2\pi}\sum_{k=1}^{\infty} \left[C^w_k \exp(-ik\phi )\right] +h.c.
    \nonumber \\
    &=  \frac{1}{2\pi} + \frac{1}{2\pi}\sum_{k=1}^{\infty} \left[A^w_k \exp\{-i(k\phi-\theta^w_k) \}\right] +h.c.
    \nonumber \\
    &=\frac{1}{2\pi} + \frac{1}{\pi}\sum_{k=1}^{\infty} A^w_k\cos(k\phi-\theta^w_k) .
    \label{eq:P(phi) W HO-1}
\end{align}
This form of $P_{w}(\phi)$ tells us that the phase distribution could be written as a superposition of different harmonic modes. Let all the coherences $\bra{m}\rho\ket{n}$ of the density matrix for a harmonic oscillator form a set S. These coherences are the resource for synchronization, without them a state produces a uniform phase distribution with a value $1/2\pi$. S could be further divided into subsets $\{S_k\}$ based on the density matrix elements of the form $\bra{m+k}\rho\ket{m}$. The elements of the subset $S_k$ contribute to the $k$-th harmonic of $P_{w}(\phi)$, which leads to a distribution with $k$ peaks between 0 to $2\pi$. Let the subset with the largest absolute value of $A_k^w$ be $S_{k_d}$. In other words, this is the dominant subset because due to its contribution, $k_d$ is the dominant harmonic in $P_{w}(\phi)$ and will lead to $k_d$ peaks. The position of peaks is given by $\phi_P=(\theta_{k_d}+2n\pi)/k_d$, where $n$ is an integer. Moreover, because of the form of $C_k^w$, there is also a possibility of destructive interference between the coherence terms to kill the contribution from a subset $S_k$. Next we extend this analysis for the two-oscillator case.  

\vspace{0.5cm}
{\it Two oscillator case.--} The two-mode Wigner function for a joint state $\rho$ of two continuous variable oscillators is defined as \cite{Lee-2}
\begin{equation}
    W(x_1, p_1, x_2, p_2) = \frac{1}{\pi^2} \int_{-\infty}^{\infty} \! dy_1\  \int_{-\infty}^{\infty}  dy_2 \ e^{2i (p_1 y_1 + p_2 y_2)}
\times \langle x_1 - y_1, x_2 - y_2|\rho|x_1+ y_1, x_2 + y_2\rangle. 
\end{equation}
Expanding $\rho$ in the joint Fock basis and defining $W_{mn}$ similar Eq.~\eqref{eq:Wig fock-1}, we can write the two-mode Wigner function as
\begin{equation}
    W(x_1, p_1, x_2, p_2) = \sum_{m_1 n_1 m_2 n_2} \langle m_1 m_2 |\rho| n_1 n_2 \rangle \, W_{m_1 n_1} (x_1, p_1) W_{m_2 n_2} (x_2, p_2).
\end{equation}
We perform the following transformation on the distribution
$$(x_1,p_1,x_2,y_2)\to (r_1,\phi_1,r_2,\phi_2)\to(r_1,r_2,\phi,\phi_2)$$
The first transformation is from cartesian to polar coordinates and in the second transformation, we substitute $\phi_1=(\phi+\phi_2)$ with $\phi$ being the relative phase. We next integrate out the variables $r_1,r_2$ and $\phi_2$ to get the marginal probability distribution $P_{w}(\phi)$ as 
\begin{align}
    P_w(\phi) =& \sum_{m_1 n_1 m_2 n_2} \langle m_1 m_2 |\rho| n_1 n_2 \rangle \int^{\infty}_0 dr_1\  r_1 \int^{\infty}_0 dr_2\  r_2 \int^{2\pi}_0 d\phi_2\ W_{m_1 n_1} W_{m_2 n_2} \nonumber \\
    =&\frac{1}{(2\pi)^2}\sum_{m_1 n_1 m_2 n_2} \langle m_2 m_2 |\rho| n_1 n_1 \rangle  r_w(m_1,n_1) r_w(m_2,n_2) \times \nonumber\\ 
    &\exp\{ i (n_1-m_1)\phi \}\int^{2\pi}_0 d\phi_2\ \exp\left\{ i[(n_1+ n_2)-(m_1+m_2)]\phi_2 \right\}.
\end{align}
Here $r_w$ is defined in a similar way as done in Eq.~\eqref{rwmn}. Importantly, the integral with respect to $\phi_2$ survives only for those density matrix elements $\bra{m_1,m_2}\rho\ket{n_1,n_2}$ terms which satisfy the following excitation relation:
\begin{equation}
    m_1 + m_2 = n_1 + n_2.
    \label{eq:selection rule QHO_supp}
\end{equation}
Here we obtain a condition over the joint density matrix elements which determine the relative phase distribution between two quantum harmonic oscillators, unlike the case of a single harmonic oscillator where all density matrix elements contributed to $P_{w}(\phi)$, as in Eq.~\eqref{eq:1 HO P(phi) coh}. The simplified form of the distribution is given by
\begin{equation}
    P_w(\phi) =\frac{1}{2\pi} \sum_{ m_1+m_2= n_1+n_2} \langle m_1 m_2 |\rho| n_1 n_2 \rangle\   R_w(m_1,n_1;m_2,n_2)\   \exp\{i(n_1 - m_1)\phi\}
    \label{eq: 2 QHO P(phi) coh}
\end{equation}
where $ R_w(m_1,n_1;m_2,n_2) = r_w(m_1,n_1) r_w(m_2,n_2)$. Because of this form: $R_w(m_1,m_1;m_2,m_2)=1$ and $ R_w(m_1,n_1;m_2,n_2)= R_w(n_1,m_1;n_2,m_2)$. As before, we define a new term $C_k^w$ as follows
\begin{equation}
     C_k^w =\sum_{\substack{ m_1+m_2= n_1+n_2\\m_1-n_1=k}}\bra{m_1,m_2}\rho\ket{n_1,n_2} R_w(m_1,n_1;m_2,n_2),
     \label{Ckw}
\end{equation}
which could be decomposed into its radial and angular parts as $C_k^w=A_k^w\exp(i\theta_k^w)$. Simplifying Eq \eqref{eq: 2 QHO P(phi) coh} further we get
\begin{align}
    P_w(\phi)&=\frac{1}{2\pi} + \frac{1}{2\pi} \sum_{\substack{ m_1+m_2= n_1+n_2\\ m_1 \neq n_1}} \langle m_1 m_2 |\rho| n_1 n_2 \rangle\    R_w(m_1,n_1;m_2,n_2)\   \exp\{i(n_1 - m_1)\phi\}\nonumber \\
    &=\frac{1}{2\pi} +  \frac{1}{2\pi} \sum_{\substack{ m_1+m_2= n_1+n_2\\ m_1 > n_1}} \langle m_1 m_2 |\rho| n_1 n_2 \rangle\    R_w(m_1,n_1;m_2,n_2)\   \exp\{i(n_1 - m_1)\phi\}+ h.c. \nonumber \\
    &=\frac{1}{2\pi} + \frac{1}{2\pi}\sum_{k=1}^{\infty} \left[ \exp(-ik\phi )\sum_{\substack{ m_1+m_2= n_1+n_2\\m_1-n_1=k}}\bra{m_1,m_2}\rho\ket{n_1,n_2} R_w(m_1,n_1;m_2,n_2)\right] +h.c.
    \nonumber \\
    &= \frac{1}{2\pi} + \frac{1}{2\pi}\sum_{k=1}^{\infty} \left[C^w_k \exp(-ik\phi )\right] +h.c.
    \nonumber \\
    &=  \frac{1}{2\pi} + \frac{1}{2\pi}\sum_{k=1}^{\infty} \left[A^w_k \exp\{-i(k\phi-\theta^w_k) \}\right] +h.c.
    \nonumber \\
    &=\frac{1}{2\pi} + \frac{1}{\pi}\sum_{k=1}^{\infty} A^w_k\cos(k\phi-\theta^w_k).
    \label{eq:P(phi) W QHO-2}
\end{align}
Therefore, the relative phase distribution can be written as a superposition of different harmonic modes. More importantly, the condition in Eq.~\eqref{eq:selection rule QHO_supp} states which elements of the joint density matrix are important for determining the relative phase distribution. These elements form a set S. This set could be further divided into subsets $S_k$ based on classification $\bra{m_1+k,m_2}\rho\ket{m_1,m_2+k}$. Elements of the subset $S_k$ contribute to the $k$-th harmonic of the relative phase distribution.

\subsection{Husimi Q-Function}
\label{sec: App Huisimi Q-Function}
{\it Single oscillator case.--} The Husimi Q-function for a harmonic oscillator with a state $\rho$ is defined as
\begin{equation}
    Q(\alpha) = \frac{1}{\pi}\bra{\alpha}\rho\ket{\alpha}
    \label{q func 1 HO}
\end{equation}
where $\ket{\alpha}$ is the coherent state for the harmonic oscillator and is defined as $\ket{\alpha} = e^{-\frac{|\alpha|^2}{2}} \sum_{n=0}^{\infty} \frac{\alpha^n}{\sqrt{n!}} \ket{n}$. Expanding the equation for Q in terms of the Fock basis elements, we get
\begin{align}
     Q(\alpha) &= \frac{1}{\pi} \sum_{m,n=0}^{\infty} \langle m | \rho | n \rangle \bra{\alpha}m\rangle\langle n \ket{\alpha} \nonumber \\
     &= \frac{1}{\pi} \sum_{m,n=0}^{\infty} \langle m | \rho | n \rangle\  \exp(-|\alpha|^2)\  \frac{\alpha^n \alpha^{*m}}{\sqrt{n!m!}},
\end{align}
where $\alpha$ is a complex number and could be written as $\alpha = re^{i\phi}$. Similar to the case of a Wigner function we integrate out $r$ to find the marginal distribution $P_q(\phi)$, where the subscript $q$ refers to the Q-distribution. We get
\begin{align}
    P_q(\phi) &= \frac{1}{\pi} \sum_{m,n=0}^{\infty}\langle m | \rho | n \rangle\ \int^{\infty}_0 dr\  r\ \exp(-r^2)\  \frac{r^{n+m}}{\sqrt{n!m!}} \ \exp\{i(n-m)\phi\} \nonumber \\
     &= \frac{1}{2\pi} \sum_{m,n=0}^{\infty} \langle m | \rho | n \rangle\  r_q(m,n)\  \exp\{i(n-m)\phi\},
     \label{eq: P(phi) Q coh}
\end{align}
where $r_q(m,n)$ is the integral of the radial part of the Q-function, given by
\begin{equation}
    r_q(m,n) = \int^{\infty}_0 dr\  2r\ \exp(-r^2)\  \frac{r^{n+m}}{\sqrt{n!m!}}\ ,
    \label{eq:R(m,n) q}
\end{equation}
with similar properties as $r_w(m,n)$: $r_q(m,m) = 1$ and $r_q(m,n) = r_q(n,m)$. We define a new term $C_k^q$ as follows
\begin{equation}
     C_k^q = \sum_{m-n=k}\bra{m}\rho\ket{n}r_q(m,n),
     \label{Ckq}
\end{equation}
which could be decomposed into its radial and angular parts as $C_k^q=A_k^q\exp(i\theta_k^q)$. Eq.~\eqref{eq: P(phi) Q coh} could be further simplified, similar to Eq.~\eqref{eq:P(phi) W HO-1}, to reduce it to a superposition of different harmonic modes. The simplified form of $P_q(\phi)$ is given by
\begin{equation}
    P_q(\phi)=\frac{1}{2\pi} + \frac{1}{\pi}\sum_{k=1}^{\infty} A^q_k\cos(k\phi-\theta^q_k).
    \label{eq:P(phi) q 1 HO}
\end{equation}
Therefore, once again we receive a similar expression like the Wigner case for the single oscillator. However, note the difference in the expression for $C_k^{q}$ given in Eq.~\eqref{Ckq} compared to the expression for $C_{k}^{w}$, as given in Eq.~\eqref{eq:ckw}. We next extend this calculation for the two-oscillator case.

\vspace{0.5cm}

{\it Two oscillator case.--} For the state $\rho$ corresponding to a pair of harmonic oscillators, the two-mode Q-function is defined as
\begin{equation}
    Q(\alpha_1,\alpha_2)= \frac{1}{\pi^2}\bra{\alpha_1,\alpha_2}\rho\ket{\alpha_1,\alpha_2}.
    \label{eq: 2 QHO Q-func}
\end{equation}
Expanding Eq.~\eqref{eq: 2 QHO Q-func} in the joint Fock basis we get
\begin{align}
   Q &=  \frac{1}{\pi^2} \sum_{m_1 n_1 m_2 n_2} \langle m_1 m_2 |\rho| n_1 n_2 \rangle \  \langle \alpha_1 \ket{m_1}\langle \alpha_2 \ket{m_2}\langle n_1 \ket{\alpha_1}\langle n_2 \ket{\alpha_2} \nonumber \\
   &= \frac{1}{\pi^2} \sum_{m_1 n_1 m_2 n_2} \langle m_1 m_2 |\rho| n_1 n_2 \rangle \  \exp(-|\alpha_1|^2)\  \frac{\alpha_1^{n_1} \alpha_1^{*{m_1}}}{\sqrt{n_1!m_1!}} \  \exp(-|\alpha_2|^2)\ \frac{\alpha_2^{n_2} \alpha_2^{*{m_2}}}{\sqrt{n_2!m_2!}}.
\end{align}
Here $\alpha_1$ and $\alpha_2$ are complex numbers and could be written as $\alpha_1 = r_1e^{i\phi_1}$ and $\alpha_2 = r_2e^{i\phi_2}$. This gives us
\begin{equation}
   Q = \frac{1}{\pi^2} \sum_{m_1 n_1 m_2 n_2} \langle m_1 m_2 |\rho| n_1 n_2 \rangle \exp(-r_1^2)\frac{r_1^{n_1+m_1}}{\sqrt{n_1!m_1!}}\exp(-r_2^2)\frac{r_2^{n_2+m_2}}{\sqrt{n_2!m_2!}}\exp\{i(n_1-m_1)\phi_1\}\exp\{i(n_2-m_2)\phi_2\}. 
\end{equation}
Now transforming the equation in terms of the relative phase $\phi$ and the variable $\phi_2$, we get
\begin{align}
   Q = \frac{1}{\pi^2} \sum_{m_1 n_1 m_2 n_2} & \langle m_1 m_2 |\rho| n_1 n_2 \rangle \  \exp(-r_1^2)\  \frac{r_1^{n_1+m_1}}{\sqrt{n_1!m_1!}} \  \exp(-r_2^2)\ \frac{r_2^{n_2+m_2}}{\sqrt{n_2!m_2!}} \nonumber \\
   & \times \exp\{ i (n_1-m_1)\phi \}\  \exp\left\{ i[(n_1+ n_2)-(m_1+m_2)]\phi_2 \right\}
\end{align}
To obtain $P_q(\phi)$ we integrate out $r_1,r_2$ and $\phi_2$ as follows
\begin{align}
   P_q(\phi) = \frac{1}{\pi^2} \sum_{m_1 n_1 m_2 n_2} & \langle m_1 m_2 |\rho| n_1 n_2 \rangle\int^{\infty}_0 dr_1\  r_1   \exp(-r_1^2)\  \frac{r_1^{n_1+m_1}}{\sqrt{n_1!m_1!}} \ \int^{\infty}_0 dr_2\  r_2\exp(-r_2^2)\ \frac{r_2^{n_2+m_2}}{\sqrt{n_2!m_2!}} \nonumber \\
   & \exp\{ i (n_1-m_1)\phi \}\int^{2\pi}_0 d\phi_2\ \exp\left\{ i[(n_1+ n_2)-(m_1+m_2)]\phi_2 \right\}.
\end{align}
Integration over $\phi_2$ gives us the same condition as in Eq.~\eqref{eq:selection rule QHO_supp}, over the density matrix elements i.e., the elements that satisfy the following excitation relation 
\begin{equation}
m_1+m_2=n_1+n_2
\end{equation} 
contributes to phase locking. This simplifies the equation to be
\begin{equation}
    P_q(\phi) =\frac{1}{2\pi} \sum_{m_1+m_2= n_1+n_2} \langle m_1 m_2 |\rho| n_1 n_2 \rangle\  R_q(m_1,m_2,n_1,n_2)\   \exp[i{(n_1 - m_1)\phi}]
    \label{eq:P(phi) coh Q 2 QHO}
\end{equation}
where $ R_q(m_1,n_1;m_2,n_2) = r_q(m_1,n_1) r_q(m_2,n_2)$, with $R_q(m_1,m_1;m_2,m_2)=1$ and $ R_q(m_1,n_1;m_2,n_2)= R_q(n_1,m_1;n_2,m_2)$ . We once again define a new complex term $C_k^q$ as follows
\begin{equation}
     C_k^q = \sum_{\substack{ m_1+m_2= n_1+n_2\\m_1-n_1=k}}\bra{m_1,m_2}\rho\ket{n_1,n_2}R_q(m_1,m_2,n_1,n_2),
\end{equation}
which could be decomposed into its angular and radial parts as $C_k^q=A_k^q\exp(i\theta_k^q)$. Simplifying Eq.~\eqref{eq:P(phi) coh Q 2 QHO} similar to Eq.~\eqref{eq:P(phi) W QHO-2},  we get
\begin{equation}
    P_q(\phi)=\frac{1}{2\pi} + \frac{1}{\pi}\sum_{k=1}^{\infty} A^q_k\cos(k\phi-\theta^q_k),
    \label{eq:P(phi) Q QHO-2}
\end{equation}
which shows that the Phase distribution expressions given in Eq.~\eqref{eq:P(phi) q 1 HO} and Eq.~\eqref{eq:P(phi) Q QHO-2}, derived from the Husimi Q-function are very similar to the ones derived from the Wigner function, with only difference being in the radial integral part.

\subsection{Harmonic oscillator Phase states}

{\it Single oscillator case--} Instead of using the Quasiprobability distributions like Wigner function and the Husimi Q distribution, we could also use harmonic oscillator phase states to plot the phase distributions. The phase states are defined as \cite{SG}
\begin{equation}
    \ket{\phi}= \sum_{n=0}^{\infty} e^{in\phi}\ket{n}.
    \label{eq:SG Eigenstates_app}
\end{equation}
Such a state is an eigenstate of Susskind-Glogower (SG) phase operator $\hat{E}\equiv(\hat{n}+1)^{-\frac{1}{2}}\hat{a}$  \cite{SG,PhaseReview}.
The Phase distribution of a single harmonic oscillator in the state $\rho$, in terms of these phase states is given by \cite{POVM}
\begin{align}
     P_p(\phi)&=\frac{1}{2\pi}\bra{\phi}\rho\ket{\phi}\nonumber \\
     &= \frac{1}{2\pi} \sum_{m,n=0}^{\infty} \langle m | \rho | n \rangle \bra{\phi}m\rangle\langle n \ket{\phi} \nonumber \\
     &= \frac{1}{2\pi} \sum_{m,n=0}^{\infty} \langle m | \rho | n \rangle\  \exp\{i(n-m)\phi\} 
     \label{eq:P(phi) ps coh}
\end{align}
We define a new complex term $C_k^p$ as \begin{equation}
     C_k^p = \sum_{m-n=k}\bra{m}\rho\ket{n},
\end{equation}
where for this phase state case, $ C_k^p$ is simply the sum of all elements of the subset $S_k$. This quantity can be decomposed into radial and angular parts as $C_k^p=A_k^p\exp(i\theta_k^p)$. $P_p(\phi)$ can be simplified to the following form
\begin{equation}
    P_p(\phi)=\frac{1}{2\pi} + \frac{1}{\pi}\sum_{k=1}^{\infty} A^p_k\cos(k\phi-\theta^p_k) .
    \label{eq:P(phi)_ps_1_HO}
\end{equation}

\vspace{0.3cm}
{\it Two oscillator case.--} For a pair of harmonic oscillators in state $\rho$, the joint phase distribution is given by \cite{BlockadeBruder}
\begin{align}
    P(\phi_1,\phi_2)&= \frac{1}{(2\pi)^2} \bra{\phi_1,\phi_2} \rho\ket{\phi_1,\phi_2} \nonumber \\
    &= \frac{1}{(2\pi)^2} \sum_{m_1 n_1 m_2 n_2} \langle m_1 m_2 |\rho| n_1 n_2 \rangle \  \langle \phi_1 \ket{m_1}\langle \phi_2 \ket{m_2}\langle n_1 \ket{\phi_1}\langle n_2 \ket{\phi_2} \nonumber \\
    &= \frac{1}{(2\pi)^2} \sum_{m_1 n_1 m_2 n_2} \langle m_1 m_2 |\rho| n_1 n_2 \rangle \exp\{i(n_1-m_1)\phi_1\} \exp\{i(n_2-m_2)\phi_2\}.
\end{align}
To find the relative phase distribution, we perform a variable transform from $(\phi_1,\phi_2)$ to $(\phi,\phi_2)$ and integrate out $\phi_2$. We obtain 
\begin{equation}
    P_p(\phi) = \frac{1}{(2\pi)^2} \sum_{m_1 n_1 m_2 n_2} \langle m_1 m_2 |\rho| n_1 n_2 \rangle  \exp\{ i (n_1-m_1)\phi \}\int^{2\pi}_0 d\phi_2\ \exp\left\{ i[(n_1+ n_2)-(m_1+m_2)]\phi_2 \right\}.
\end{equation}
In this case also, we get the same excitation relation
\begin{equation}
m_1+m_2=n_1+n_2
\end{equation} 
involving the density matrix elements as in Eq.~\eqref{eq:selection rule QHO_supp} because of the integration over $\phi_2$. Using this condition, we get
\begin{equation}
    P_q(\phi) =\frac{1}{2\pi} \sum_{m_1+m_2= n_1+n_2} \langle m_1 m_2 |\rho| n_1 n_2 \rangle \ \exp\{i{(n_1 - m_1)\phi}\}.
    \label{eq:P(phi) coh PS 2 QHO}
\end{equation}
We define a new term $C_k^p$ as follows
\begin{equation}
     C_k^p =\sum_{\substack{ m_1+m_2= n_1+n_2\\m_1-n_1=k}}\bra{m_1,m_2}\rho\ket{n_1,n_2},
\end{equation}
which could be decomposed into its radial and angular parts as $C_k^p=A_k^p\exp(i\theta_k^p)$. We simplify form for $P_p(\phi)$ similar to Eq.~\eqref{eq:P(phi) W QHO-2} and \eqref{eq:P(phi) Q QHO-2}. We finally obtain the phase distribution as
\begin{equation}
    P_p(\phi)=\frac{1}{2\pi} + \frac{1}{\pi}\sum_{k=1}^{\infty} A^p_k\cos(k\phi-\theta^p_k).
    \label{eq:P(phi)_P_QHO-2}
\end{equation}
Note that the phase distributions in Eq.~\eqref{eq:P(phi)_ps_1_HO} and Eq.~\eqref{eq:P(phi)_P_QHO-2}, derived from phase states $\ket{\phi}$ are just dependent on density matrix elements, unlike the distributions derived from the phase space approach (Wigner and Husimi) which also had contributions from radial integrals. This form helps us better understand the relation between a given state (density matrix) and the corresponding phase.

\section{Phase distribution of a Spin}
\label{app:spin}
We next move to the discrete variable i.e., for the spin case. In this section, we provide detailed derivations for the phase distributions for spin systems.

\subsection{Husimi Q-Function}
{\it Single spin case--} The phase space distribution of a single spin could be derived from the Husimi Q function, which in this case is defined in terms of the spin coherent states $\ket{\theta,\phi}$. For a spin-$s$, the spin coherent state $\ket{\theta,\phi}$ is defined by rotating the maximal state $\ket{s,m=s}$ counterclockwise first by an angle $\theta$ about the $y$-axis, and then by an angle $\phi$ about the $z$-axis \cite{LYMAJP_2015,1-Spin}
\begin{equation}
    | \theta, \phi \rangle = \exp(-i \phi S_z) \exp(-i \theta S_y) | s, s \rangle,
    \label{SCS-supp}
\end{equation}
where the spin operators $S_i$, $i=x, y, z$ are the generators of the rotation group SO(3) satisfying $[S_j,S_k]=i\epsilon_{jkl}S_l$ and $\pi \leq \theta \leq 0$ and $ 0 \leq \phi < 2 \pi$. Due to such a rotation of the extremal state in Eq.~\eqref{SCS-supp}, we get $\langle \theta, \phi| \bf{\hat{S}}|\theta \phi\rangle = \bf{s}$, where $\textbf{s}= (s \sin \theta \cos \phi, s \sin \theta \sin \phi, s \cos \theta)$.
Using the Wigner-D matrix \cite{WignerD, LYMAJP_2015}, these spin coherent states can be written as
\begin{align}
    |\theta, \phi\rangle &= \sum_{m=-s}^{s} \binom{2s}{s+m} ^{\frac{1}{2}} \Big(\cos \frac{\theta}{2} \Big)^{s+m} \Big(\sin \frac{\theta}{2}\Big)^{s-m} e^{-im\phi} |s, m\rangle \nonumber \\
    &=\sum_{m=-s}^{s} N_{s,m}(\theta) e^{-im\phi} |s, m\rangle,
  \label{eq:spin coherent}
\end{align}
where we define $N_{s,m}(\theta)$ as 
\begin{equation}
    N_{s,m}(\theta) = \binom{2s}{s+m} ^{\frac{1}{2}}  \Big(\cos \frac{\theta}{2} \Big)^{s+m} \Big(\sin \frac{\theta}{2}\Big)^{s-m}.
\end{equation}
For a single spin $s$, and described by an arbitrary density matrix $\rho$, the associated normalized Husimi Q function is defined as
\begin{equation}
  Q(\theta, \phi) = \frac{2s + 1}{4\pi} \langle \theta, \phi | \rho | \theta, \phi \rangle.
  \label{eq:spin Q}
\end{equation}
We integrate out the $\theta$ variable to obtain the marginal distribution for $P_{q}(\phi)$ as
\begin{align}
    P_q(\phi)&=\int_0^{\pi}d\theta \sin \theta \, Q(\theta,\phi)
    \label{eq:P(phi) spin -1} \nonumber \\
    &=\frac{2s + 1}{4\pi}\int_0^{\pi}d\theta\sin \theta \, \langle \theta, \phi | \rho | \theta, \phi \rangle.
\end{align}
Expanding the density matrix in the basis $\{\ket{s,m}\}$, we obtain
\begin{align}
    P_q(\phi)&=\frac{2s + 1}{4\pi}\sum_{m,n=-s}^{s} \bra{s,m}\rho\ket{s,n}\int_0^{\pi}d\theta\sin(\theta)  \langle \theta, \phi \ket{s,m} \bra{s,n}\theta, \phi \rangle, \nonumber \\
    &=\frac{2s + 1}{4\pi}\sum_{m,n=-s}^{s}\bra{s,m}\rho\ket{s,n}\int_0^{\pi}d\theta\sin(\theta)  N_{s,m}(\theta)N_{s,n}(\theta)\exp\{i(m-n)\phi\}, \nonumber \\
     &=\frac{2s + 1}{4\pi}\sum_{m,n=-s}^{s} \bra{s,m}\rho\ket{s,n} t_q(m,n)\exp\{i(m-n)\phi\},
     \label{eq:1 spin P(phi) coh}
\end{align}
where $t_q(m,n)$ involves the integral over the $\theta$ variable is given as
\begin{equation}
    t_q(m,n) = \int_0^{\pi}d\theta\sin(\theta)  N_{s,m}(\theta)N_{s,n}(\theta),
\end{equation}
with properties $t_q(m,m)=2/(2s+1)$ and $t_q(m,n)=t_q(n,m)$. We next define a new complex term $C_k^q$ as follows
\begin{equation}
     C_k^q = \sum_{m-n=k}\bra{s,m}\rho\ket{s,n}t_q(m,n),
\end{equation}
which could be decomposed into its radial and angular part as $C_k^q=A_k^q\exp(i\theta_k^q)$. Using this definition, Eq.~\eqref{eq:1 spin P(phi) coh} could be further expanded as
\begin{align}
    P_q(\phi)&=\frac{2s + 1}{4\pi}\sum_{mn}\bra{s,m}\rho\ket{s,n}t_q(m,n)\exp\{i(m-n)\phi\}\nonumber \\
    &=\frac{1}{2\pi}+\frac{2s + 1}{4\pi}\sum_{m\neq n}\bra{s,m}\rho\ket{s,n}t_q(m,n)\exp\{i(m-n)\phi\}\nonumber \\
    &=\frac{1}{2\pi} + \frac{2s + 1}{4\pi} \sum_{m > n} \langle s,m | \rho | s,n \rangle  t_q(m,n)\  \exp\{i(m-n)\phi \} + \langle s,n | \rho | s,m \rangle  t_q(n,m)\  \exp\{i(n-m)\phi \}\nonumber \\
    &=\frac{1}{2\pi} + \frac{2s + 1}{4\pi}\sum_{k=1}^{2s} \left[ \exp(ik\phi )\sum_{m-n=k}\bra{s,m}\rho\ket{s,n}t_q(m,n)\right] +h.c.
    \nonumber \\
    &= \frac{1}{2\pi} + \frac{2s + 1}{4\pi}\sum_{k=1}^{2s} \left[C^q_k \exp(ik\phi )\right] +h.c.
    \nonumber \\
    &=  \frac{1}{2\pi} + \frac{2s + 1}{4\pi}\sum_{k=1}^{2s} \left[A^q_k \exp\{i(k\phi+\theta^q_k) \}\right] +h.c.
    \nonumber \\
    &=\frac{1}{2\pi} + \frac{2s + 1}{2\pi}\sum_{k=1}^{2s} A^q_k\cos(k\phi+\theta^q_k)\ .
    \label{eq:P(phi) spin-1 Q}
\end{align}
As can be seen, similar to the case of the harmonic oscillator, the phase distribution has been reduced to a superposition of different harmonic modes. All the density matrix elements which contribute to phase locking, form a set S, these elements could be further divided into subsets $S_k$ with elements $\bra{s,m+k}\rho\ket{s,m}$. The elements of $S_k$ contribute to the $k$-th mode of the distribution. The subset $S_k$ with the largest $A_q^k$ is the dominant subset, we label it as $k_d$. This means that $k_d$ harmonic mode dominates the phase distribution and its major features will be determined by $\cos(k_d\phi+\theta^q_{k_d})$. The distribution will have $k_d$ peaks between 0 to $2\pi$, with the location given by $\phi_p=(2n\pi-\theta^q_{k_d})/k_d$, where $n$ is an integer.

\vspace{0.3cm}
{\it Two spin case--} For two spins $s_1$ and $s_2$, the two-mode Husimi Q function for the joint state $\rho$ is defined as
\begin{equation}
Q(\theta_1, \phi_1, \theta_2, \phi_2) = \frac{(2s_1 + 1)(2s_2 + 1)}{(4\pi)^2} \langle \theta_1, \phi_1 ; \theta_2, \phi_2 | \rho | \theta_2, \phi_2 ;\theta_1, \phi_1 \rangle, 
\label{eq:two mode q function}
\end{equation}
where $\ket{\theta_i,\phi_i}$ is the spin coherent state for the $i$-th spin. Expanding $\rho$ in the joint basis $\{\ket{s_1,m_1;s_2,m_2}\}$ we get,
\begin{align}
     Q=  \frac{(2s_1 + 1)(2s_2 + 1)}{(4\pi)^2} \sum_{m_1,m_2,n_1,n_2} & \bra{s_1,m_1,;s_2,m_2}\rho\ket{s_1,n_1;s_2,n_2} N_{s_1,m_1}(\theta_1) N_{s_2,m_2}(\theta_2)N_{s_1,n_1}(\theta_1)N_{s_2,n_2}(\theta_2) \nonumber \\
     & \exp\{i(m_1 - n_1)\phi_1\}   \exp\{i(m_2 - n_2)\phi_2\}.
     \label{eq:2 spin Q func}
 \end{align}
Here note that the variables $m_i, n_i$, $i=1,2$ takes value between $-s_i$ and $s_i$. We next perform a variable transform of the joint Q function from $(\theta_1,\theta_2,\phi_1,\phi_2)$ to $(\theta_1,\theta_2,\phi,\phi_2)$, where $\phi = (\phi_1 - \phi_2)$, and then integrate out the $\theta_1,\theta_2$ and $\phi_2$ variables to get a distribution in the relative phase $\phi$. We get
\begin{equation}
    P_q(\phi) = \int_0^{\pi} d\theta_1 \sin \theta_1  \int_0^{\pi} d\theta_2 \sin \theta_2  \int_0^{2\pi} d\phi_2\  Q(\theta_1,\theta_2,\phi,\phi_2) 
\end{equation}
Plugging in the explicit equation for the Q function given in Eq.~\eqref{eq:2 spin Q func}, we get
\begin{align}
    P_q(\phi)= \frac{(2s_1 + 1)(2s_2 + 1)}{(4\pi)^2} & \sum_{m_1,m_2,n_1,n_2}  \bra{s_1,m_1;s_2,m_2}\rho\ket{s_1,n_1;s_2,n_2}  \exp\{i(m_1-n_1)\phi\} 
     \nonumber \\ 
    & \int_0^{\pi} d\theta_1 \sin(\theta_1) N_{s_1,m_1}(\theta_1)N_{s_1,n_1}(\theta_1) \int_0^{\pi} d\theta_2 \sin(\theta_2) N_{s_2,m_2}(\theta_2)N_{s_2,n_2}(\theta_2) \times \nonumber\\
    & \int_0^{2\pi} d\phi_2\ \exp\{i\left[(m_1+m_2)-(n_1+n_2)\right]\phi_2\}
    \label{eq:2_spin_integrate}
\end{align}
Similar to the case of relative phase distribution for two harmonic oscillators, the integral over $\phi_2$ eliminates those density matrix terms that {\it do not} satisfy the condition $m_1+m_2 = n_1 + n_2$. This gives us exactly the same excitation relation for phase locking as the case for two harmonic oscillators. In other words, the relative phase distribution between two arbitrary spins is determined by those joint density matrix elements $\bra{s_1,m_1;s_2,m_2}\rho\ket{s_1,n_1;s_2,n_2}$ which satisfy the excitation relation 
\begin{equation}
    m_1+m_2 = n_1 + n_2.
    \label{eq:selection rule spin}
\end{equation}
Using the excitation relation, Eq.~\eqref{eq:2_spin_integrate} simplifies to
\begin{align}
    P_q(\phi) = \frac{(2s_1 + 1)(2s_2 + 1)}{8\pi} \sum_{m_1+m_2=n_1+n_2} &\bra{s_1,m_1;s_2,m_2}\rho\ket{s_1,n_1;s_2,n_2} \times \nonumber \\
    &T_q(m_1,n_1;m_2,n_2)\, \exp\{i(m_1-n_1)\phi\},
\end{align}
where $T_q(m_1,n_1;m_2,n_2)=t_q(m_1,n_1)t_q(m_2,n_2)$ with properties $T_q(m_1,m_1;m_2,m_2)=4/[(2s_1+1)(2s_2+1)]$ and $T_q(m_1,n_1;m_2,n_2)=T_q(n_1,m_1;n_2,m_2)$. We define a new complex term $C_k^q$ as follows
\begin{equation}
     C_k^q =\sum_{\substack{ m_1+m_2= n_1+n_2\\m_1-n_1=k}}\bra{s_1,m_1;s_2,m_2}\rho\ket{s_1,n_1;s_2,n_2}T_q(m_1,n_1;m_2,n_2),
\end{equation}
which could be decomposed into its radial and angular parts as $C_k^q=A_k^q\exp(i\theta_k^q)$. The relative phase distribution could be reduced to a superposition of different harmonic modes as follows
\begin{equation}
    P_q(\phi)=\frac{1}{2\pi} + \frac{(2s_1 + 1)(2s_2 + 1)}{4\pi}\sum_{k=1}^{\min\{2s_1,2s_2\}} A^q_k\cos(k\phi+\theta^q_k)\ .
    \label{eq:P(phi) q 2 spin}
\end{equation}

The elements following the excitation relation in Eq.~\eqref{eq:selection rule spin}, which contribute to the relative phase distribution form a set S. This set could be further divided into subsets $S_k$ based on the classification $\bra{s_1,m_1+k;s_2,m_2}\rho\ket{s_1,m_1;s_2,m_2+k}$. Elements of the subset $S_k$ contribute to the $k$-th harmonic of the relative phase distribution. These results are quite similar to the harmonic oscillator case.

\subsection{Spin Phase States}
Similar to the case of the harmonic oscillator, one could also use states with well-defined phase for the spins to find the phase distribution. For a spin when we talk about phase $\phi$, in the classical sense, this is the angle subtended by spin vector $\mathbf{\hat{S}}$ on the $x$-$y$ plane. Studies on the phase operator $\hat{\phi}$ for a spin and its exponential $e^{i\hat{\phi}}$ have previously been done \cite{Spin_phase_states}. For a spin-s system, eigenstates for phase operator $\hat{\phi}$ are given by,
\begin{equation}
    |\phi_n\rangle = \frac{1}{\sqrt{2s + 1}} \sum_{m=-s}^{s} e^{-im\phi_n} |s,m\rangle.
\end{equation}
The phase eigenvalues $\{\phi_n\}$ for a spin are discretized, with $\phi_n = \frac{2n\pi}{2s+1}$. In the classical limit for the spin when $s\to\infty$,  $\{\phi_n\}$ becomes continuous over $(0,2\pi)$. The (2s+1) phase eigenstates $\{\ket{\phi_n}\}$ form a complete orthonormal basis for the spin Hilbert space. To get a continuous distribution over the phase, similar to the case of quasiprobability distribution, we allow $\phi_n$ to take continuous values over the range $(0,2\pi)$. As the phases are not discretized anymore, we simply write $\phi_n\equiv\phi$ and $\ket{\phi_n}\equiv\ket{\phi}$. The states $\{\ket{\phi}\}$ now form an overcomplete basis for the Hilbert space, similar to the spin coherent states, yielding a resolution to unity, $ \frac{2s+1}{2\pi} \int_{0}^{2\pi} d\phi \, \ket{\phi}\bra{\phi} = 1$. The phase distribution for a single spin-s in terms of the phase state is given by
\begin{equation}
    P(\phi)=\frac{2s+1}{2\pi}\bra{\phi}\rho\ket{\phi}
\end{equation}
where
\begin{equation}
|\phi\rangle = \frac{1}{\sqrt{2s + 1}} \sum_{m=-s}^{s} e^{-im\phi} |s,m\rangle.
\end{equation}
We define a new complex term $C_k^p$ as follows
\begin{equation}
     C_k^p = \sum_{m-n=k}\bra{s,m}\rho\ket{s,n},
\end{equation}
which could be decomposed into a radial and an angular part as $C_k^p=A_k^p\exp(i\theta_k^p)$. The above equation could be simplified to a form similar to Eq.~\eqref{eq:P(phi) spin-1 Q}, the final relative phase distribution could be written as
\begin{equation}
    P_p(\phi)=\frac{1}{2\pi} + \frac{1}{\pi}\sum_{k=1}^{2s} A^p_k\cos(k\phi+\theta^p_k).
    \label{eq:P(phi) ps 1 spin}
\end{equation}
For a system of two spins, $s_1$ and $s_2$, the joint phase distribution is given by:
\begin{equation}
    P(\phi_1,\phi_2)= \frac{(2s_1+1)(2s_2+1)}{(2\pi)^2} \bra{\phi_1,\phi_2} \rho\ket{\phi_1,\phi_2}.
\end{equation}
Similar to previous case in Eq \ref{eq:2_spin_integrate}, transforming the variables $\phi_1$ and $\phi_2$ to $\phi$ and $\phi_2$, then integrating over $\phi_2$ to get $P_p(\phi)$, we obtain the same excitation relation as in Eq.~\eqref{eq:selection rule spin}. We define a new term $C_k^p$ as follows
\begin{equation}
     C_k^p = \sum_{\substack{ m_1+m_2= n_1+n_2\\m_1-n_1=k}}\bra{s_1,m_1;s_2,m_2}\rho\ket{s_1,n_1;s_2,n_2},
\end{equation}
it could be decomposed into its angular and radial parts as $C_k^p=A_k^p\exp(i\theta_k^p)$. The relative phase distribution obtained from phase states could be simplified to a form similar to Eq \ref{eq:P(phi) q 2 spin} as follows
\begin{equation}
    P_p(\phi)=\frac{1}{2\pi} + \frac{1}{\pi}\sum_{k=1}^{\min\{2s_1,2s_2\}} A^p_k\cos(k\phi+\theta^p_k)\ .
    \label{eq:P(phi) q 2 spin}
\end{equation}
The distributions Eq.~\eqref{eq:P(phi) ps 1 spin} and Eq.~\eqref{eq:P(phi) q 2 spin} derived from phase states are only dependent on density matrix elements, unlike the ones derived from phase space distribution, which are also dependent on theta integrals.

\section{Phase Distribution for a Hybrid System}
\label{sec:AppendixC}
In the previous two sections, we discussed phase synchronization either between two quantum harmonic oscillators or two spin systems. Given that both a quantum harmonic oscillator and a spin have their own defined phases, we now discuss about phase locking for a hybrid setup consisting of a single spin of magnitude $s$ and a single harmonic oscillator.

\subsection{Husimi Q Function for the hybrid setup}
We consider a joint system made out of a quantum harmonic oscillator and a spin. Let $\rho$ be the joint density matrix, representing an arbitrary state. The joint Husimi Q-function is then defined as
\begin{equation}
Q(r, \phi_o, \theta_s, \phi_s) = \frac{2s+1}{4\pi^2} \bra{\alpha_o,\alpha_s} \rho \ket{\alpha_o,\alpha_s},
\end{equation}
whereby $\ket{\alpha_o}$, we denote the coherent state for a quantum harmonic oscillator, and $\ket{\alpha_s} \equiv |\theta_s, \phi_s\rangle$ is the spin coherent state for the spin $s$. As seen in previous sections, the $r$ and the $\theta$ variables determine the populations for the oscillator and the spin, and in contrast, $\phi_o$ and $\phi_s$ are the desired phases for the oscillator and the spin systems. The basis for the oscillator is the usual Fock states, denoted here as ${\ket{m_o}}$, whereas the basis for the spin is chosen as the eigenstates for $\hat{S}_z$, i.e. ${\ket{s,m_s}}$. As our setup consists of a single spin, we write $\ket{s,m_s} \equiv \ket{m_s}$. Expanding the density matrix in this joint basis, we receive
\begin{equation}
    Q(r, \phi_o, \theta, \phi_s) = \frac{2s+1}{4\pi^2} \sum_{m_o,m_s,n_o,n_s} \bra{m_o,m_s}\rho\ket{n_o,n_s} \langle \alpha_o, \alpha_s\ket{m_o,m_s}\bra{n_o,n_s}\alpha_o, \alpha_s\rangle,
    \label{joint-Q}
\end{equation}
where note that the variables $m_0, n_0$, representing the oscillator range, from $0$ to $\infty$, whereas the variables $m_s, n_s$ representing the spins, range from $-s$ to $+s$. 
Using the form of the coherent states for the oscillator and for the spin, we get
\begin{equation}
   \langle n_o\ket{\alpha_o} = \frac{\exp\{-r^2/2\}}{\sqrt{n_o!}}\   r^{n_o}\   \exp\{i n_o\phi_o\}\ ;\ \langle n_s\ket{\alpha_s} = N_{n_s}(\theta)\   \exp\{-i n_s\phi_s\}. 
   \label{proj}
\end{equation}
Here we expressed $\alpha_0=r \, e^{i \phi_o}$, and for the spin part we use Eq.~\eqref{eq:spin coherent}. Plugging Eq.~\eqref{proj} in Eq.~\eqref{joint-Q}, the Husimi Q function simplifies to 
\begin{align}
    Q(r, \phi_o, \theta, \phi_s) = \frac{2s+1}{4\pi^2} \sum_{m_o,m_s,n_o,n_s} & \bra{m_o,m_s}\rho\ket{n_o,n_s} 
    \frac{1}{\sqrt{m_o!n_o!}} \exp\{-r^2\} \ r^{m_o+n_o} \ \exp\{i(n_o-m_o)\phi_o\}  \, \times \nonumber \\
    & N_{m_s}(\theta)\, N_{n_s}(\theta) \, \exp\{i(m_s-n_s)\phi_s\}.
\end{align}
Defining the relative phase between the oscillator and the spin as $\phi = (\phi_o - \phi_s)$ and performing transformation of the variables from $(\phi_o,\phi_s)$ to $(\phi,\phi_s)$, we get
\begin{align}
    Q(r, \theta, \phi, \phi_s) = \frac{2s+1}{4\pi^2} \sum_{m_o,m_s,n_o,n_s}& \bra{m_o,m_s}\rho\ket{n_o,n_s}\ \frac{1}{\sqrt{m_o!n_o!}}   \exp\{-r^2\} \ r^{m_o+n_o} \   N_{m_s}(\theta)N_{n_s}(\theta) \nonumber \\
    & \exp\{i(n_o-m_o)\phi\} \exp\{i [(n_o - m_o) + (m_s - n_s) ]\phi_s\}.
\end{align}
Now we integrate out the $r$, $\theta$ and $\phi_s$ variables to obtain the final form of the relative phase distribution $P_q(\phi)$ as
\begin{align}
    P_q(\phi) = \frac{2s+1}{8\pi^2} \sum_{m_o,m_s,n_o,n_s}& \bra{m_o,m_s}\rho\ket{n_o,n_s}\   \int_0^{\infty} dr\ \frac{1}{\sqrt{m_o!n_o!}} \exp\{-r^2\} \ 2r^{m_o+n_o+1} \   \int_0^{\pi} d\theta\   \sin(\theta)\   N_{m_s}(\theta)N_{n_s}(\theta) \nonumber \\
    & \exp\{i(n_o - m_o)\phi\} \int_0^{2\pi}d\phi_s \,  \exp\{i[(m_s - n_s)-(m_o-n_o) ]\phi_s\}. 
\end{align}
Similar to the earlier cases, the integration of $\phi_s$ leads to elimination of certain terms from the equation for $P_q(\phi)$ and only those joint density matrix elements $\bra{m_o,m_s}\rho\ket{n_o,n_s}$ determine the relative phase distribution which satisfy the following excitation relation
\begin{equation}
    m_o - n_o = m_s - n_s.
    \label{eq:selection rule hybrid}
\end{equation}
Note the sign difference in the hybrid case, in comparison to the two-oscillator or two-spin cases. This simplifies the equation for $P_q(\phi)$ to
\begin{equation}
    P_q(\phi) = \frac{2s+1}{4\pi} \sum_{m_o - n_0 = m_s - n_s} \bra{m_o,m_s}\rho\ket{n_o,n_s} I_q(m_o,n_o;m_s,n_s) \exp\{i(n_o - m_o)\phi\},
\end{equation}
where $I_q(m_o,n_o;m_s,n_s) = r_q(m_o,n_o)t_s(m_s,n_s)$ with properties $I_q(m_o,m_o;m_s,m_s)=2/(2s+1)$ and $I_q(m_o,n_o;m_s,n_s)=I_q(n_o,m_o;n_s,m_s)$. Using these properties, we re-write the equation for $P_q(\phi)$ in terms of diagonal and off-diagonal terms as
\begin{equation}
    P_q(\phi) =\frac{1}{2\pi} + \frac{2s+1}{4\pi}\sum_{\substack{ m_o - n_0 = m_s - n_s\\m_o\neq n_o}}\bra{m_o,m_s}\rho\ket{n_o,n_s} I_q(m_o,n_o;m_s,n_s) \exp\{i(n_o - m_o)\phi\}. 
\end{equation}
All the joint density matrix elements $\bra{m_o,m_s}\rho\ket{n_o,n_s}$ satisfying the excitation relation given in Eq.~\eqref{eq:selection rule hybrid} form a set $S$, which could be further divided into subsets $S_k$ based on the classification $(m_o-n_o)=k$. A general element of subset $S_k$ could be written as $\bra{m_o+k,m_s+k}\rho\ket{m_o,m_s}$. We define a new term $C_k^q$ as follows
\begin{equation}
     C_k^q = \sum_{\substack{ m_o - n_0 = m_s - n_s\\m_o-n_o=k}}\bra{m_o,m_s}\rho\ket{n_o,n_s} I_q(m_o,n_o;m_s,n_s),
\end{equation}
which could be decomposed into its radial and angular parts as $C_k^q=A_k^q\exp(i\theta_k^q)$. Similar to previous cases, the distribution could be reduced to a superposition of different harmonic modes as follows
\begin{equation}
    P_q(\phi)=\frac{1}{2\pi} + \frac{2s + 1}{2\pi}\sum_{k=1}^{2s} A^q_k\cos(k\phi-\theta^q_k)
\end{equation}

\subsection{Phase States for the hybrid setup}
 
The joint phase distribution for the hybrid system using spin and harmonic oscillator phase states is
\begin{equation}
     P(\phi_o,\phi_s)= \frac{2s+1}{(2\pi)^2} \bra{\phi_o,\phi_s} \rho\ket{\phi_o,\phi_s}.
\end{equation}
Expanding $\rho$ in the joint basis $\ket{m_o,m_s}$ we get 
\begin{align}
    P(\phi_o,\phi_s)&= \frac{2s+1}{(2\pi)^2} \sum_{m_o n_o m_s n_s} \langle m_o m_s |\rho| n_o n_s \rangle \  \langle \phi_o \ket{m_o}\langle \phi_s \ket{m_s}\langle n_o \ket{\phi_o}\langle n_s \ket{\phi_s} \nonumber \\
    &= \frac{1}{(2\pi)^2} \sum_{m_o n_o m_s n_s} \langle m_o m_s |\rho| n_o n_s \rangle \exp\{i(n_o-m_o)\phi_o\} \exp\{i(m_s-n_s)\phi_s\}.
\end{align}
Performing transformation of the variables from $(\phi_o,\phi_s)$ to $(\phi,\phi_s)$ and integrating out $\phi_s$ we get
\begin{equation}
    P_p(\phi) =\frac{1}{(2\pi)^2} \sum_{m_o n_o m_s n_s} \langle m_o m_s |\rho| n_o n_s \rangle \exp\{i(n_o-m_o)\phi\}\int^{2\pi}_0 d\phi_s\ \exp\left\{ i[(m_s- n_s) -(m_o - n_o)]\phi_s \right\}.
\end{equation}
In this case too, we get the same excitation relation for the density matrix elements as in Eq.~\eqref{eq:selection rule hybrid}. Using the excitation relation we get
\begin{equation}
    P_p(\phi) =\frac{1}{2\pi} \sum_{m_o-m_s= n_o-n_s} \langle m_o m_s |\rho| n_o n_s \rangle \ \exp\{i{(n_o - m_o)\phi}\}.
\end{equation}
We define a new term $C_k^p$ as follows
\begin{equation}
     C_k^p =\sum_{\substack{ m_o-m_s= n_o-n_s\\m_o-n_o=k}}\bra{m_o,m_s}\rho\ket{n_o,n_s},
\end{equation}
which could be decomposed into its radial and angular parts as $C_k^p=A_k^p\exp(i\theta_k^p)$. Using these terms, the final form of $P_p(\phi)$ could be written as
\begin{equation}
    P_p(\phi)=\frac{1}{2\pi} + \frac{1}{\pi}\sum_{k=1}^{2s} A^p_k\cos(k\phi-\theta^p_k).
    \label{eq:P(phi) P Hybrid}
\end{equation}

\section{Perturbative Calculations for continuous variable systems}
\label{sec:AppendixD}

In this section, we use the perturbation theory to analytically understand the generation of specific coherence elements due to interactions between two continuous variable oscillators \cite{Anharmonic,Quadratically}. For example, consider a system of two uncoupled quantum VdP oscillators with master equation
\begin{equation}
\dot{\rho} = -i \left[ H_0, \rho \right] + \sum_{i=1,2} \gamma_g^{(i)} \mathcal{D}\left[a_i^\dagger \right] \rho + \gamma_l^{(i)} \mathcal{D}\left[a_i^2 \right] \rho \equiv \mathcal{L}_0 \rho.
\label{eq:QME_two_VdP_supp}
\end{equation}
Any small interaction between the two VdPs can be treated as a perturbation to the Liouvillian $\mathcal{L}_0$, with a small strength $\epsilon$. This gives us the final Liouvillian to be of the form 
\begin{equation}
    \mathcal{L}=\mathcal{L}_0+\epsilon\mathcal{L}_I.
\end{equation}
Now we perform a perturbative expansion of the steady state as follows:
\begin{equation}
    \rho_{ss} =  \rho_{ss}^{(0)}+ \epsilon\rho_{ss}^{(1)}+\epsilon^2\rho_{ss}^{(2)}+\ ...
\end{equation}
where $\rho_{ss}^{(0)}$ is the joint steady state of the two VdP oscillators under $\mathcal{L}_0$. We further have $\tr(\rho_{ss}^{(i)})=0$, for $i>0$. Under the steady state condition, we must have $\mathcal{L}\rho_{ss}=0$. Expanding this equation we get the $n$-th order correction to be
\begin{equation}
    \mathcal{L}_0\rho_{ss}^{(n)}=-\mathcal{L}_I\rho_{ss}^{(n-1)}.
\end{equation}
Using the Moore-Penrose pseudoinverse $\mathcal{L}_0^{-1}$, we could invert the action of $ \mathcal{L}_0$ and write \cite{Perturb,Sai_Coh_Degen}
\begin{equation}
    \rho_{ss}^{(n)}=-\mathcal{L}_0^{-1}\mathcal{L}_I\rho_{ss}^{(n-1)},
\end{equation}
where note that $\mathcal{L}_I$ is the term which creates coherences resulting in phase synchronization. Action of $\mathcal{L}_0$ on the operator $\ket{m_1+k,m_2}\bra{m_1,m_2+k}$, which results in an element of subset $S_k$, creates a superposition of following operators 
\begin{align}
    &\ket{m_1+k,m_2}\bra{m_1,m_2+k} \nonumber \\
    &\ket{(m_1+1)+k,m_2}\bra{m_1+1,m_2+k} \nonumber \\
    & \ket{m_1+k,m_2+1}\bra{m_1,(m_2+1)+k} \nonumber \\
    &\ket{(m_1-2)+k,m_2}\bra{m_1-2,m_2+k} \nonumber \\
    &\ket{m_1+k,m_2-2}\bra{m_1,(m_2-2)+k},
\end{align}
all of which lead to elements within the same subset $S_k$. This shows that the action of $\mathcal{L}_0$ on a state $\rho$ couples elements within the same subset $S_k$. If a certain subset has all elements zero then it will remain so under the action of ${\cal L}_0$. We claim that the action of $\mathcal{L}_0^{-1}$ is similar. To see this, consider a state $\rho_k$ that has non-zero matrix elements only within the subset $S_k$, while all other elements are zero, and satisfying the condition $\tr(\rho_k)=0$. Now, consider the following equation
\begin{equation}
\mathcal{L}_0\mathcal{L}_0^{-1}\rho_k=\rho_k, 
\end{equation}
if the action of $\mathcal{L}_0^{-1}$ creates coherences outside the subset $S_k$, then $\mathcal{L}_0$ cannot bring it back inside the subset which means the final result cannot be $\rho_k$, which leads to a contradiction. This means that even $\mathcal{L}_0^{-1}$ couples elements only within the same subset $S_k$. So $\mathcal{L}_I$ creates coherences and $\mathcal{L}_0$ modifies these coherences by creating superpositions within the subsets. 

\subsection{Synchronizing Interactions}

{\it Coherent Coupling -- }For coherent exchange interaction of the form $V=(a_1^{\dagger}a_2+a_1a_2^{\dagger})$, the interaction Liouvillian is given by
\begin{equation}
    \mathcal{L}_I\rho = -i[a_1^{\dagger}a_2+a_1a_2^{\dagger},\rho].
\end{equation}
Action of $\mathcal{L}_I$ on $\ket{m_1+k,m_2}\bra{m_1,m_2+k}$ creates a superposition of following operators
\begin{align}
    &\ket{(m_1-1)+(k+1),m_2}\bra{m_1-1,m_2+(k+1)} \nonumber \\
    & \ket{(m_1+1)+(k-1),m_2}\bra{m_1+1,m_2+(k-1)} \nonumber \\
    &\ket{m_1+(k+1),m_2-1}\bra{m_1,(m_2-1)+(k+1)} \nonumber \\
    &\ket{m_1+(k-1),m_2+1}\bra{m_1,(m_2+1)+(k-1)}.  
    \label{eqn:coh_terms_perturb}
\end{align}
This gives us coherences within subsets $S_{k+1}$ and $S_{k-1}$. Moreover, these coherences are within the same total occupation subspaces as $\ket{m_1+k,m_2}\bra{m_1,m_2+k}$, with $N_T= (m_1+m_2+k)$, this is expected as interaction is excitation preserving. As $\rho_{ss}^{(0)}$ is the diagonal matrix, i.e. with elements of $S_0$, $\rho_{ss}^{(1)}$ will have coherences of the subset $S_1$ and $S_{-1}$ (which is simply the complex conjugate of $S_1$). In  $\rho_{ss}^{(2)}$ we will have diagonal correction and coherences of subset $S_2$ (and its complex conjugate). Higher order corrections to will have coherences from higher subsets. This is how coherent coupling generates coherences following the excitation relation in Eq.~\eqref{eq:selection rule QHO_supp}, which leads to phase locking between the two VdPs.

\vspace{0.5cm}
{\it Dissipative Coupling -- } For the correlated one photon loss of the form $\mathcal{D}[a_1+a_2]$, the interaction Liouvillian is given by
\begin{equation}
    \mathcal{L}_I\rho = \mathcal{D}[a_1+a_2]\rho
\end{equation}
Action of $\mathcal{L}_I$ on $\ket{m_1+k,m_2}\bra{m_1,m_2+k}$ creates a superposition of terms which include all the elements of Eq.~\eqref{eqn:coh_terms_perturb} and also the additional two elements
\begin{align}
    &\ket{m_1+(k-1),m_2}\bra{m_1,m_2+(k-1)} \nonumber \\
    &\ket{(m_1-1)+(k+1),m_2-1}\bra{m_1-1,(m_2-1)+(k+1)}
\end{align}
this gives us coherences within subsets $S_{k+1}$ and $S_{k-1}$. Moreover, for this case apart from having coherences within the same total occupation subspace $(N_T= m_1+m_2+k)$, they also exist for subspaces with $(N_T= m_1+m_2+k-1)$, which is expected as the coupling is dissipative. Similar to the coherent case, higher-order corrections give coherences from higher subsets. These generated coherences lead to phase locking between the oscillators. 

\subsection{Non-synchronizing Interaction}

{\it Two-Mode Squeezing -- }For two-mode squeezing interaction of the form $V=i(a_1a_2-a_1^{\dagger}a_2^{\dagger})$, the interaction Liouvillian is given by
\begin{equation}
    \mathcal{L}_I\rho = -i[i(a_1a_2-a_1^{\dagger}a_2^{\dagger}),\rho].
\end{equation}
Action of $\mathcal{L}_I$ on $\ket{m_1+k,m_2}\bra{m_1,m_2+k}$ creates a superposition of following operators
\begin{align}
    &\ket{m_1+k,m_2}\bra{m_1-1,m_2+k-1} \nonumber \\
    & \ket{m_1+k,m_2}\bra{m_1+1,m_2+k+1} \nonumber \\
    &\ket{m_1+k+1,m_2+1}\bra{m_1,m_2+k} \nonumber \\
    &\ket{m_1+k-1,m_2-1}\bra{m_1,m_2+k}.  
\end{align}
This gives us coherences which do not follow the excitation relation and therefore do not result in synchronization. 

\vspace{0.5cm}

{\it Dissipative Coupling -- } Similarly, for the case of dissipative coupling of the form $\mathcal{D}[a_1+a_2^2]$, the action of $\mathcal{L}_I$ doesn't generate elements following the excitation relation of Eq.~\eqref{eq:selection rule QHO_supp}, that is why it leads to no synchronization.

\vspace{0.5cm}

This perturbative approach can be naturally extended to spin systems and hybrid setups to investigate the impact of various interaction forms on coupled self-sustained oscillators. By analyzing how different interactions perturb the steady state of the uncoupled system, one can identify the specific types of coherence generated and assess whether they satisfy the excitation relation relevant to each system. Such an extension enables a systematic exploration of which interactions are capable of inducing synchronization in a given physical setting.

\twocolumngrid
\bibliography{references}

\end{document}